\documentclass[10pt]{article}
\usepackage[T1]{fontenc}
\usepackage{amssymb}
\usepackage{slashed}
\usepackage{cancel}
\usepackage{mathrsfs} %for mathscr fonts
\usepackage{fullpage}
\usepackage{mathtools} %needed before hyperref in order to avoid bugs in links to eqns
\usepackage[colorlinks=true,linkcolor=blue,citecolor=magenta,linktocpage=true]{hyperref}
\usepackage[numbers,sort,compress]{natbib}
\usepackage{tocloft}
\usepackage{titlesec}
\usepackage{units}
\usepackage{tikz}
\usepackage{tikz-cd}
\usepackage{amsthm}
\newtheorem{proposition}{Proposition}

\usepackage{empheq}
\usepackage[most]{tcolorbox}
\newtcbox{\othermathbox}[1][]{nobeforeafter,math upper,tcbox raise base,enhanced,rounded corners,colback=black!5,colframe=black}
\def\be#1\ee{\begin{align}#1\end{align}}
\def\bsub#1\esub{\begin{subequations}#1\end{subequations}}

\def\q{\qquad}
\def\f{\frac}

\def\tf{\text{\tiny{TF}}}

\def\de{\mathrm{d}}

\def\C{\mathcal{C}}
\def\D{\mathcal{D}}

\def\F{\mathcal{F}}

\def\J{\mathcal{J}}
\def\K{\mathcal{K}}

\def\M{\mathcal{M}}
\def\N{\mathcal{N}}
\def\O{\mathcal{O}}
\def\P{\mathcal{P}}

\def\S{\mathcal{S}}

\def\ra{\rangle}
\def\la{\langle}

\newcommand{\rsemidirsum}{%
  \mathbin{\!%
    \begin{tikzpicture}[baseline=(plus.base), scale=0.12]
      \node (plus) at (0,0) {$+$};

      \draw[line width=0.55pt]  % thickened
        ([xshift=-1.05]plus.center) ++(0,1.05)
          arc[start angle=90, end angle=-90, radius=1.05];
    \end{tikzpicture}%
 \! }%
}

\numberwithin{equation}{section}
\def\d{\text{d}}
\begin{document}

\title{\Large{\textbf{\sffamily Lifting asymptotic symmetries\\from metric to tetrad formulation}}}
\author{\sffamily Antoine Vincenti$^1$}
\date{}
\date{\small{\textit{
${}^1$ENS de Lyon, CNRS, LPENSL, UMR 5672, 69342 Lyon cedex 07, France}}}

\maketitle

%%%%%%%%%%%%%%%%%%%%%%%%%%%%%%%%%%%%%%%%
%%%%%%%%%%%%%%%%%%%%%%%%%%%%%%%%%%%%%%%%

\begin{abstract}
Metric solution spaces defined by gauge and boundary conditions are associated with a residual symmetry algebra, consisting of the diffeomorphism generators whose action preserves these conditions. Describing a metric solution space in terms of tetrad variables adds a local Lorentz freedom. When part of this freedom is fixed by additional tetrad gauge and boundary conditions, metric residual symmetries must be supplemented by compensating Lorentz transformations. If a residual Lorentz freedom remains, there is an ambiguity in these compensating transformations. The most direct choices generally fail to realize the metric residual symmetry algebra as a subalgebra of the tetrad one, because of a field-dependent cocycle. We show that a reference tetrad, that is a section of the tetrad solution space seen as a principal bundle over the metric one, always provides a lift free of this cocycle, so that the tetrad residual symmetry algebra is the semi-direct sum of the metric one and the ideal of field-dependent residual Lorentz transformations. The construction is independent of the gauge conditions, of the residual group, of the first-order Lagrangian, and reduces to the Kosmann lift on metric isometry generators. We illustrate it on a Weyl BMS-type solution space with time-dependent boundary metric, described in tetrad variables with a residual U(1) Lorentz freedom, for which the cocycle of a naive lift is computed and then explicitly removed.
\end{abstract}

%%%%%%%%%%%%%%%%%%%%%%%%%%%%%%%%%%%%%%%%
%%%%%%%%%%%%%%%%%%%%%%%%%%%%%%%%%%%%%%%%

\thispagestyle{empty}

\newpage
\setcounter{page}{1}
\hfill
\tableofcontents
\bigskip
\hrule

%%%%%%%%%%%%%%%%%%%%%%%%%%%%%%%%%%%%%%%%
%%%%%%%%%%%%%%%%%%%%%%%%%%%%%%%%%%%%%%%%

\newpage

\section{Introduction}
In general relativity, asymptotic symmetries are commonly described, in the gauge-fixing approach, as the residual diffeomorphism preserving a choice of gauge and boundary conditions defining a space of solutions; see \textit{e.g.} \cite{Ruzziconi:2019pzd} for a review. Although asymptotic symmetries can also, in many situations, be defined in a coordinate-independent way through conformal compactification \cite{Penrose2011, Geroch:1977big}, we follow here this coordinate-based approach, where gauge conditions are imposed in the bulk. The seminal Bondi–van der Burg–Metzner–Sachs analysis of asymptotically flat spacetimes at null infinity and their famous BMS asymptotic symmetries \cite{Bondi:1962px, Sachs:1962wk, Sachs:1962zza} belongs to this approach and has known a renewed interest in the past decades. This has led in particular to various extensions of the BMS symmetries, obtained by relaxing the boundary conditions historically considered \cite{Barnich:2009se, Barnich:2010eb, Campiglia:2014yka, Freidel:2021fxf, Geiller:2022vto}, but also to the analysis of other asymptotic regimes, for instance at spatial infinity \cite{Henneaux:2018cst}, with a non-vanishing cosmological constant \cite{Compere:2019bua}, at finite-distance boundaries \cite{Chandrasekaran:2018aop, Freidel:2020xyx} or with cosmological asymptotics \cite{Bonga:2020fhx, Campoleoni:2026abr}.

%A characteristic feature of the residual symmetries appearing in these various works is that their generators generically present field-dependence. Their algebra is therefore computed with the modified bracket \cite{Barnich:2007bf, Barnich:2010eb}. What is then usually called the asymptotic symmetry algebra is the parameter algebra of the generators, which closes with field-independent structure constants only in favorable cases, possibly after a field-dependent redefinition of the parameters.}

These analyses are mostly carried out in metric variables but a first-order description, in terms of tetrad variables, is needed or motivated in many contexts, \textit{e.g.} to couple fermions \cite{Aneesh:2020fcr, Prabhu:2021bod}, to use the Newman–Penrose formalism \cite{Newman:1961qr, Barnich:2019vzx}, or to include the Holst, Euler and Pontryagin terms in the action \cite{Holst:1995pc, Corichi:2016zac}. The latter leave the field equations unchanged but modify the symplectic structure and therefore the charges that one can associate with residual symmetries \cite{DePaoli:2018erh, Oliveri:2019gvm}, giving for instance dual gravitational charges \cite{Godazgar:2018qpq, Godazgar:2020kqd, Oliveri:2020xls}.

In tetrad variables, the gauge symmetries are described locally as diffeomorphisms together with local Lorentz transformations. In a tetrad description of a metric solution space, one usually fixes at least part of the local Lorentz freedom by imposing additional tetrad gauge conditions. In order to preserve them, the action of metric residual symmetries on the tetrad must be supplemented by compensating Lorentz transformations \cite{Godazgar:2020kqd, Freidel:2021fxf}. When the additional tetrad gauge conditions leave a residual Lorentz freedom, these compensating transformations are only defined up to a residual Lorentz transformation. The natural choices, obtained by solving explicitly the preservation of tetrad gauge conditions, generally fail to provide a realization of the metric residual symmetries as a subalgebra of the tetrad ones: their bracket produces an additional residual Lorentz transformation, given by a field-dependent cocycle. Beyond the algebra, this is a problem at the level of charges associated with residual symmetries. Indeed, through such a cocycle, residual Lorentz charges, if non-vanishing, appear in the metric charge algebra, which then no longer represent the metric residual symmetry algebra. This raises two questions. In a tetrad description of a metric solution space with a partial Lorentz gauge fixing, can the compensating Lorentz transformations be chosen so that the metric residual symmetry algebra survives as a subalgebra of the tetrad one? And what is then the structure of the full tetrad residual symmetry algebra?

In this paper, we give a general answer to these questions, which to our knowledge have not been addressed systematically. We show that the answer to the first question is always positive, the right choice being provided by a reference tetrad, that is, the assignment to each metric of the given solution space of a tetrad satisfying the chosen gauge conditions. More precisely, consider a generic metric solution space $\S_g$. We describe it in tetrad variables and impose additional tetrad gauge conditions which are preserved by local Lorentz transformations valued in a residual subgroup $H$ of the Lorentz group. The resulting tetrad solution space, denoted by $\S_e^H$, has the structure of a principal bundle over the metric one. A reference tetrad is then simply a section of this field-space bundle. The projectable tetrad residual symmetries, whose vector part depends only on the metric, form an algebra $\mathfrak{a}_\pi^H$ which extends the metric residual algebra $\mathfrak{g}$ by the ideal of field-dependent residual Lorentz transformations. A reference tetrad always provides a lift of $\mathfrak{g}$ into $\mathfrak{a}_\pi^H$ which is a Lie algebra homomorphism, free of any cocycle, so that we get the semi-direct sum structure 
\be
\label{eq:structure}
\mathfrak{a}_\pi^H=\mathfrak{g}\rsemidirsum\C^\infty(\S^H_e,\mathfrak{l}_H),
\ee
where $\mathfrak{l}_H$ denotes the local Lorentz generators valued in the Lie algebra of $H$. When the residual Lorentz transformations are taken field-independent, this semi-direct sum reduces to a direct sum, so that in particular a parameterization of the metric residual symmetries with field-independent structure constants always induces one of the tetrad residual symmetries. The result does not depend on the precise form of the gauge conditions, on the residual group, which needs neither to be abelian nor normal, or on the additional covariant terms that can be added to the Lagrangian. It provides a practical shortcut to compute the lift without solving the preservation of the tetrad gauge conditions. On generators of spacetime isometries, the lift reduces to the Kosmann lift, which underlies the Lorentz-diffeomorphism charges used in black hole thermodynamics in \textit{e.g.} \cite{Jacobson:2015uqa, Prabhu:2015vua, Aneesh:2020fcr}; it can thus be seen as its extension to asymptotic symmetries. 

We finally illustrate this construction on the concrete example of the metric BMSW solution space with time-dependent boundary metric, described in tetrad variables with a residual local U(1) freedom. At the level of the solution space and the asymptotic symmetry algebra, this generalizes the analysis of \cite{Freidel:2021fxf, Godazgar:2022foc}. We show how a naive lift of the metric asymptotic symmetry algebra can be obtained in that case and how a non-vanishing and field-dependent cocycle then arises in the tetrad algebra. We then show how this cocycle can be removed by hand through a field-dependent redefinition of the residual U(1) generators, and finally how these computations can be shortcut by a direct use of the general results we found.

The paper is organized as follows. In section \ref{sec:firstorder}, we recall some aspects of first-order formalism, the notions of field-dependent generator and modified bracket, then we define the residual symmetry algebra associated with a solution space. Section \ref{sec:3} contains the general construction outlined above, which ends up on the structure result of equation \eqref{eq:structure}, followed by a discussion of its upshots. Section \ref{sec:BMSW-u-U(1)} treats the example in detail, and section \ref{sec:perspectives} gives some perspectives for this work. A discussion of the global geometry of first-order formalism and of the notion of covariant lifts can be found in appendix \ref{app:geometry}, and appendix \ref{app:complements} gives additional material and explicit formulas for the example of section \ref{sec:BMSW-u-U(1)}.

%%%%%%%%%%%%%%%%%%%%%%%%%%%%%%%%%%%%%%%%
%%%%%%%%%%%%%%%%%%%%%%%%%%%%%%%%%%%%%%%%

\section{Gauge and residual symmetries in tetrad variables}
\label{sec:firstorder}

%%%%%%%%%%%%%%%%%%%%%%%%%%%%%%%%%%%%%%%%

\subsection{Tetrad variables and gauge symmetries}
Let us first recall some basic aspects of first-order formalism. A more detailed account of these aspects can be found in appendix \ref{app:geometry}, in particular of the global geometry framework and of the passage from global to local. For a more general review, see \textit{e.g.} \cite{Corichi:2016zac}.

The dynamical fields in first-order formalism are the tetrad $e$ and the spin connection $\omega$. In what follows, we only work locally. In the metric, second-order formulation, it means working on a coordinate patch $U$ of an underlying spacetime differential manifold $\M$, giving its indices to the metric tensor. In first-order formalism, it means working in a trivialization of an underlying Lorentz principal bundle $\P\rightarrow\M$ over the coordinate patch $U$. This allows us to write the tetrad as a one-form valued in a fixed internal vector space with Lorentzian metric $\eta_{IJ}$, with its usual indices $e^I_\mu$, and the spin connection as a one-form valued in the Lorentz Lie algebra, with its usual indices $\omega_\mu^{IJ}$.

The group acting naturally (by pullback) on local metric fields is the group Diff($U$) of diffeomorphisms of the coordinate patch. At the infinitesimal level, these diffeomorphisms are generated by vector fields $\xi\in\mathfrak{X}(U)$. These generators form a Lie algebra for the standard Lie bracket and they act on metric fields \textit{via} the Lie derivative. In first-order formalism, fields are naturally acted on by the automorphism group Aut($\P$) (see appendix \ref{app:geometry} or \textit{e.g.} \cite{Prabhu:2015vua}). The infinitesimal automorphism generators can be written locally as pairs $(\xi,\lambda)$, where $\xi\in\mathfrak{X}(U)$ is a local diffeomorphism generator and $\lambda\in\mathfrak{l}\coloneqq\C^{\infty}(U,\mathfrak{so}(3,1))$ is a local Lorentz transformation generator. The set of automorphism generators forms a Lie algebra with semi-direct sum structure $\mathfrak{aut}(\P_{\vert U})=\mathfrak{X}(U)\rsemidirsum\mathfrak{l}$, the bracket being explicitly given by
\be
\label{eq:generatorbracket}
[(\xi_1,\lambda_1),(\xi_2,\lambda_2)]=([\xi_1,\xi_2],\xi_1[\lambda_2]-\xi_2[\lambda_1]+[\lambda_1,\lambda_2]).
\ee
Using the sign conventions of \textit{e.g.} \cite{DePaoli:2018erh}, generators act on fields according to
\be
\label{eq:transformationlaws}
\delta_{(\xi,\lambda)}e_\mu^I&=\pounds_\xi e_\mu^I+{\lambda^I}_Je^J_\mu,\\
\delta_{(\xi,\lambda)}\omega_\mu^{IJ}&=\pounds_\xi\omega_\mu^{IJ}-\partial_\mu\lambda^{IJ}-[\omega_\mu,\lambda]^{IJ},
\ee
from which we can also compute their action on any field functional;  for instance, the metric $g_{\mu\nu}[e]=e^I_\mu e^J_\nu\eta_{IJ}$ induced by the tetrad, on which the action of generators is
\be
\delta_{(\xi,\lambda)}g_{\mu\nu}[e]=\pounds_\xi g_{\mu\nu}[e],
\ee
which is of course insensitive to the Lorentz part.

As in the metric case, these transformations are of importance not only because they act naturally on the fields, but because the dynamics of these fields is covariant under this action. The dynamical Lagrangian under consideration here is the Cartan--Palatini Lagrangian $L_C\propto\varepsilon_{IJKL}e^I\wedge e^J\wedge F[\omega]^{KL}$ \cite{Corichi:2016zac}. It is Lorentz invariant and diffeomorphism covariant in the sense that
\be
\label{eq:Lagrangiancovariance}
\delta_{(\xi,\lambda)}L_C=\pounds_\xi L_C=\d(\iota_\xi L_C).
\ee
Consequently, the equations of motion are covariant and the finite transformations map solutions to solutions. For that they are the gauge symmetries of the theory. Note that the non-dynamical (at least in vacuum) Holst Lagrangian, as well as the topological Euler and Pontryagin Lagrangians \cite{Holst:1995pc, Corichi:2016zac}, which can be added to $L_C$ in order to enrich the symplectic structure of the theory (see \textit{e.g.} \cite{Corichi:2016zac, Freidel:2020xyx, DePaoli:2018erh, Oliveri:2019gvm}), also satisfy \eqref{eq:Lagrangiancovariance}. 

%%%%%%%%%%%%%%%%%%%%%%%%%%%%%%%%%%%%%%%%

\subsection{Field-dependent generators and modified bracket}
As is well known from the asymptotic symmetry literature and as we will see below, computing residual symmetries of a given set of field configurations generically requires introducing field-dependent gauge symmetry generators, see \textit{e.g.} the reviews \cite{Ruzziconi:2019pzd, Compere:2018aar} and references therein. Let us recall this notion and that of modified bracket.

We denote by $\F$ the set of all field configurations, without imposing any equations of motion, gauge or boundary conditions, leaving this for the next subsection. In the following, we will discuss both metric and tetrad solution spaces, so let us write generically the gauge symmetry algebra as $\mathfrak{s}$, which can consist of diffeomorphism or automorphism generators depending on the case. A field-dependent generator is then simply a map $X : \F \rightarrow \mathfrak{s}$. Just like field-independent generators, they act as derivations on field functionals. Consequently, any field-dependent generator $X$ defines a vector field $\delta_X$ on $\F$. In particular, it acts on any other field-dependent generator. The relevant notion of bracket between two field-dependent generators $X_1$ and $X_2$ is the so-called \textit{modified bracket} \cite{Barnich:2010eb, Barnich:2010xq}, defined as
\be
[X_1,X_2]_\star\coloneqq[X_1,X_2]-\delta_{X_1}X_2+\delta_{X_2}X_1.
\ee
Let us specify this formula for the metric and tetrad cases explicitly.
\begin{itemize}
\item If $\mathfrak{s}=\mathfrak{X}(U)$, then the $X_i=\xi_i$ are field-dependent vector fields and the modified bracket is
\be
\label{eq:vectmodifiedbracket}
[\xi_1,\xi_2]_\star\coloneqq[\xi_1,\xi_2]-\delta_{\xi_1}\xi_2+\delta_{\xi_2}\xi_1.
\ee
\item If $\mathfrak{s}=\mathfrak{aut}(\P_{\vert U})$; then generators are pairs $X_1=(\xi_1,\lambda_1)$, $X_2=(\xi_2,\lambda_2)$ and the modified bracket is 
\be
[(\xi_1,\lambda_1),(\xi_2,\lambda_2)]_\star\coloneqq(\xi_{12},\lambda_{12}),
\ee
the right-hand side being given by
\bsub
\label{eq:modifiedbracket}
\be
\label{eq:metricmodifiedbracket}
\xi_{12}&\coloneqq[\xi_1,\xi_2]-\delta_{X_1}\xi_2+\delta_{X_2}\xi_1,\\
\lambda_{12}&\coloneqq\xi_1[\lambda_2]-\xi_2[\lambda_1]+[\lambda_1,\lambda_2]-\delta_{X_1}\lambda_2+\delta_{X_2}\lambda_1.
\ee
\esub
When generators are field-independent, we recover the bracket \eqref{eq:generatorbracket}. Note that the vector part may in principle also transform under the Lorentz part here.
\end{itemize}

Let us consider again two field-dependent generators $X_1$ and $X_2$, as well as their corresponding field-space vector fields $\delta_{X_1}$, $\delta_{X_2}$. For a generic field functional $\Phi$ on $\F$, the Leibniz rule gives that $\delta_{X_1}(\delta_{X_2}\Phi)$ splits into a term including the action of $X_1$ on $X_2$ and one including its action on $\Phi$. As a consequence, one can prove that the commutator of field-space vector fields associated with field-dependent generators acts according to
\be
\label{eq:representationtheorem}
[\delta_{X_1},\delta_{X_2}]=-\delta_{[X_1,X_2]_\star}.
\ee
We will refer to this useful result in the following as the \textit{representation theorem}\footnote{This result is not to be confused with the fact that charges associated with residual symmetries may represent them as a Poisson algebra on field space, discussed in the same references.}. It can be easily checked by a direct computation in the explicit cases discussed here, see \textit{e.g.} \cite{Barnich:2007bf, Henneaux:1992ig, Barnich:2000zw} for the general setting and proof. Let us point out without more details that the mathematical structure present here is that of a Lie algebroid of $\mathfrak{s}$ over $\F$, with anchor $\rho : X\mapsto-\delta_X$ \cite{Mackenzie2005GeneralTO, Kotov:2016lpx, Barnich:2010xq}. We will not use this language here and simply refer to the Lie algebra structure of field-dependent generators under the modified bracket, together with the representation theorem.

%%%%%%%%%%%%%%%%%%%%%%%%%%%%%%%%%%%%%%%%

\subsection{Solution spaces and residual symmetries}
\label{sec:solutionspaces}
Let us now discuss the notions of solution spaces and associated residual symmetries (again, for a general review, see \textit{e.g.} \cite{Ruzziconi:2019pzd}). We again keep the discussion generic here: we denote field configurations by $\phi$ and the gauge symmetry algebra by $\mathfrak{s}$, the specification to the metric or tetrad case being immediate. We define generically a solution space as a subspace of that of all field configurations by
\be
\S\coloneqq\big\{ \phi\,\,\vert\,\, E[\phi]=0, \,c^i[\phi]=0\big \},
\ee
where $E$ is the equation of motion for $\phi$ and $c^i$ stands for a collection of gauge and boundary conditions\footnote{We assume here that boundary conditions can be written in this form. For instance a fall-off condition on a parameter $f$ which admits a power expansion $\sum_{k}f_kr^{-k}$,  like $f=\O(r^{-n})$, is understood as $c^k[f]=f_k=0$ for $k<n$.} constraining $\phi$.

At any given point $\phi\in\S$, we are interested in gauge symmetry generators that are tangent to $\S$ at $\phi$, in the sense that they preserve the conditions defining $\S$ at this point. As the field equations are covariant, only the conditions $c^i$ really have to be preserved. Hence we define the set of tangent generators at $\phi$ as
\be
\mathfrak{r}_\phi\coloneqq\big\{ X\in\mathfrak{s},\,\vert\,\, \delta_Xc^i[\phi]=0\big\}.
\ee
Then, a residual symmetry generator is defined as a field-dependent generator on $\S$, everywhere tangent to $\S$ in the previous sense. The set of all residual symmetry generators of $\S$ is then
\be
\mathfrak{r}_\S\coloneqq\big\{X : \S\rightarrow\mathfrak{s}\,\vert\,X[\phi]\in\mathfrak{r}_\phi\,,\forall \phi\in\S \big\}
\ee

A key property of $\mathfrak{r}_\S$ is that it is stable under the previously defined modified bracket. This is a consequence of the representation theorem. Indeed, if $X_1$, $X_2$ are in $\mathfrak{r}_\S$, then at each point $\phi$,
\be
[\delta_{X_1},\delta_{X_2}]c^i[\phi]=0=-\delta_{[X_1,X_2]_\star}c^i[\phi],
\ee
so that $[X_1,X_2]_\star\in\mathfrak{r}_\S$. Endowed with the modified bracket, $\mathfrak{r}_\S$ is the residual symmetry algebra of $\S$.

Note that in concrete cases, one works with a \textit{parameterization} of $\mathfrak r_{\mathcal S}$. Indeed, when solving $\delta_Xc^i[\phi]=0$, one usually faces a set of differential equations that can be solved once for a generic $\phi\in\mathcal S$. When this resolution can be carried out uniformly in $\phi$, its solutions are parameterized as $X[\phi;\alpha]$, with $\alpha$ taking values in a parameter space $P$ independent of $\phi$, and then $\mathfrak r_\phi=\{X[\phi;\alpha]\mid\alpha\in P\}$. Residual symmetry generators can of course have field-dependent parameters: any of them can be written as a map $\phi\mapsto X[\phi;\alpha[\phi]]$. For field-independent parameters $\alpha_1$ and $\alpha_2$, however, we generically have
\be
\big[X[\phi;\alpha_1],X[\phi;\alpha_2]\big]_\star=X\big[\phi;C_\phi[\alpha_1,\alpha_2]\big],
\ee
where $C_\phi$ are the structure functions of the parameterization, which in general depend on $\phi$. 
In certain cases, the structure functions are independent of $\phi$, either directly or after a field-dependent redefinition of the parameters $\alpha\mapsto\alpha'=\Psi_\phi[\alpha]$. The generators with field-independent parameters then form a Lie subalgebra of $\mathfrak r_{\mathcal S}$, isomorphic to the parameter space $P$ endowed with the bracket $C$, and every residual symmetry generator is a combination of them with field-dependent coefficients. This Lie algebra of parameters is what is usually called the residual symmetry algebra. As opposed to $\mathfrak{r}_\S$, this Lie algebra is not canonical and depends on the chosen parameterization since a field-dependent redefinition of the parameters generally makes the structure functions field-dependent.

The most famous case where we have a parameter Lie algebra is the standard BMS solution space \cite{Bondi:1962px, Sachs:1962zza, Sachs:1962wk, Barnich:2011mi}. With the usual notations, the parameters are pairs $(T,Y)$ of a supertranslation generator and a conformal Killing vector of the round sphere, and they form the $\mathfrak{bms}_4$ algebra. Instances where a field-dependent parameter redefinition is performed in order to get a genuine Lie algebra are \textit{e.g.} \cite{Geiller:2021vpg, Ciambelli:2024vhy, Geiller:2025dqe}. In other cases, no parameterization with field-independent structure functions is known, and the residual symmetries are described as a Lie algebroid. An instance of this is the $\Lambda$-BMS solution space \cite{Compere:2019bua, Compere:2020lrt}.

In these terms, the main result of the present paper is an existence result for such convenient redefinitions, concerning residual Lorentz transformations rather than diffeomorphisms.

%%%%%%%%%%%%%%%%%%%%%%%%%%%%%%%%%%%%%%%%
%%%%%%%%%%%%%%%%%%%%%%%%%%%%%%%%%%%%%%%%

\section{Lifting metric residual symmetries}
\label{sec:3}

%%%%%%%%%%%%%%%%%%%%%%%%%%%%%%%%%%%%%%%%

\subsection{Tetrad description of a metric solution space}
Suppose that we work with a given metric solution space $\S_g$. As stated at the end of the last section, it is defined as the set of all the on-shell vacuum metrics satisfying certain gauge and boundary conditions $c^i_g[g]=0$ over a patch $U$ with coordinates $x^\mu$:
\be
\S_g\coloneqq\big\{g_{\mu\nu}\mid R_{\mu\nu}[g]=0, \, c^i_g[g]=0\big\}
\ee
In the next section, the metric solution space under consideration will be that of on-shell metrics in Bondi gauge with appropriate additional fall-off conditions. Again, if we define $\mathfrak{g}_g\coloneqq\{\xi\in\mathfrak{X}(U)\,\vert\,\delta_{\xi}c^i_g[g]=0\}$ the set of vector fields generating transformations that are tangent to $\S_g$ at $g$, then the residual symmetry algebra of $\S_g$ is defined as
\be
\mathfrak{g}\coloneqq\big\{\xi:\S_g\rightarrow\mathfrak{X}(U)\,\vert\,\xi[g]\in\mathfrak{g}_g\,,\forall g\in\S_g\big\},
\ee
endowed with the modified bracket \eqref{eq:vectmodifiedbracket}.

We now want to describe this solution space with tetrad variables. Since we consider on-shell configurations of the tetrad variables, the spin connection is determined by the torsion constraint in terms of the tetrad by $\omega_\mu^{IJ}[e]=e^{\alpha I}\nabla_\mu e^J_\alpha$. For this reason, the tetrad solution space can be parameterized by tetrad fields only. Without any further constraint, the tetrad solution space that we want to consider is then
\be
\S_e\coloneqq\{e^I_\mu\mid g_{\mu\nu}[e]\in\S_g\}.
\ee
Because the gauge and boundary conditions $c^i_g$ defining $\S_g$ are defined only in terms of $g_{\mu\nu}$, they are invariant under local Lorentz transformations once written in terms of $e^I_\mu$. For this reason, for any given $g_{\mu\nu}\in\S_g$, any tetrad reproducing $g_{\mu\nu}$ is in $\S_e$. The set of such tetrads carries a free and transitive action of the group of local Lorentz transformations $L\coloneqq\C^\infty(U,\text{SO}(3,1))$\footnote{As is customary, by $\text{SO}(3,1)$ we always mean its identity component $\text{SO}^+(3,1)$, and we restrict to tetrads compatible with a fixed orientation and time orientation; see appendix \ref{app:geometry}.}. If we denote by $\pi_g$ the projection of tetrads onto their associated metric, that is $\pi_g[e]\coloneqq g_{\mu\nu}[e]$, then the structure $\S_e\xrightarrow{\pi_g}\S_g$ is formally that of a principal $L$-bundle\footnote{For other examples where a principal bundle structure is defined on field space, see \textit{e.g.} \cite{Gomes:2016mwl, franccois2021note}.}. 

The next task is to study the residual symmetry algebra $\mathfrak{a}$ of $\S_e$. Let us consider $e\in\S_e$ and ask what is the set $\mathfrak{a}_e$ of gauge symmetry generators inducing field-space transformations that are tangent to $\S_e$ at $e$. These are the pairs $(\xi,\lambda)$ such that $\delta_{(\xi,\lambda)}c^i[g[e]]=0$. This equation does not constrain the Lorentz part, which can be an arbitrary element of $\mathfrak{l}$. However, it constrains the vector field part to be in the set $\mathfrak{g}_{g[e]}$ defined just above, so that we finally have $\mathfrak{a}_e=\mathfrak{g}_{g[e]}\oplus\mathfrak{l}$. The residual symmetry algebra of $\S_e$ is then
\be
\mathfrak{a}\coloneqq\big\{(\xi,\lambda):\S_e\rightarrow\mathfrak{aut}(\P_{\vert U})\,\vert\,\xi[e]\in\mathfrak{g}_{g[e]}\,,\lambda[e]\in\mathfrak{l}\,,\forall e\in\S_e\big\},
\ee
endowed with the modified bracket \eqref{eq:modifiedbracket}. 

Among residual symmetry generators of $\S_e$, some have the property of being \textit{projectable} onto residual symmetry generators of $\S_g$. These are the ones for which the vector field part only depends on $e$ through $g[e]$, that is the generators $(\xi,\lambda)$ such that $\xi=\xi'\circ\pi_g$ for a unique $\xi'\in\mathfrak{g}$. The projection map $p:(\xi,\lambda)\mapsto\xi'$ is then well-defined, hence the name. These generators form a set
\be
\mathfrak{a}_\pi\coloneqq\big\{(\xi,\lambda)\in\mathfrak{a}\,\vert\,\xi=\xi'\circ\pi_g\,,\xi'\in\mathfrak{g}\big\}.
\ee
 As we want to see how the tetrad residual symmetry algebra can realize the metric one, we will concentrate now on this set\footnote{Also, non-projectable generators do not induce well-defined transformations of the metric solution space, see \ref{sec:summ}.}. What is the structure of $\mathfrak{a}_\pi$? First, we can see that if $(\xi_1,\lambda_1)$ and $(\xi_2,\lambda_2)$ are in $\mathfrak{a}_\pi$, the vector part of their bracket does not get any dependence on the tetrad $e$ at which it is computed except through $g[e]$, meaning that the whole bracket still lies in $\mathfrak{a}_\pi$. This implies that $\mathfrak{a}_\pi$ is a valid subalgebra of $\mathfrak{a}$. Then, elements of the form $(\xi,0)=(\xi'\circ\pi_g,0)$ with $\xi'\in\mathfrak{g}$ are in $\mathfrak{a}_\pi$ and their bracket does not produce any Lorentz part, so they are also of this form. In fact, we have an injective Lie algebra homomorphism $\xi\in\mathfrak{g}\mapsto(\xi\circ\pi_g,0)\in\mathfrak{a}_\pi$. Finally, pure Lorentz generators, that is elements of the form $(0,\lambda)$, are also in $\mathfrak{a}_\pi$. Their bracket with any other element of $\mathfrak{a}_\pi$ does not produce a vector field part, thanks to the projectability hypotheses. It means that they define an ideal $\C^\infty(\S_e,\mathfrak{l})$\footnote{We of course identify here $\lambda\in\mathfrak{l}$ with $(0,\lambda)$, and smoothness is just formal as the dimension is infinite here.}. Note that this ideal is the kernel of the projection map $p$. Altogether, what all this implies is the following structure result:
\be
\label{eq:structureofapi}
\mathfrak{a}_\pi=\mathfrak{g}\rsemidirsum\C^\infty(\S_e,\mathfrak{l}),
\ee
the semi-direct action of $\mathfrak{g}$ on $\C^\infty(\S_e,\mathfrak{l})$ being given by $[(\xi,0),(0,\lambda)]_\star=(0,\xi[\lambda]-\delta_\xi\lambda)$.

Let us recall that this result is obtained by asking nothing of the tetrads of $\S_e$ other than reproducing the metrics of $\S_g$. The realization of $\mathfrak{g}$ as a subalgebra of $\mathfrak{a}_\pi$ is easily obtained in that case, and the goal of the rest of the section is to see how to obtain this realization and a similar structure result when part of the Lorentz freedom is fixed.

%%%%%%%%%%%%%%%%%%%%%%%%%%%%%%%%%%%%%%%%

\subsection{Partial Lorentz gauge fixing and the lift problem}
\subsubsection*{Partially gauge-fixed solution space}
For concrete calculations, one may want to reduce the Lorentz freedom by imposing convenient gauge and boundary conditions on the tetrad field itself, see \textit{e.g.} \cite{Godazgar:2020kqd, Godazgar:2020gqd, Freidel:2021fxf} or \cite{Barnich:2019vzx, Mao:2024jpt, Geiller:2025dqe} for a Newman--Penrose context. Let us collectively denote these conditions by $c^j_e[e]=0$. We assume that they are preserved exactly under local Lorentz transformations with values in a subgroup $H$ of the Lorentz group, such that for any $e$  with $c^j_e[e]=0$,
\be
\label{eq:H-invariance}
c^j_e[\Lambda e]=0 \Leftrightarrow \Lambda\in L_H\subset C^\infty(U,H),
\ee
where $L_H$ is the group of local $H$-valued Lorentz transformations that are, in addition, compatible with the boundary conditions that the $c^j_e$ may include.

We now consider the subspace $\S_e^H$ of the solution space $\S_e$ studied above, made of tetrads satisfying these additional tetrad gauge conditions, that is
\be
\S_e^H\coloneqq\big\{e^I_\mu\mid e\in\S_e,\, c^j_e[e]=0\big\},
\ee
and we assume further that every metric in $\S_g$ is reproduced by at least one tetrad of $\S_e^H$. By the assumption \eqref{eq:H-invariance}, two tetrads of $\mathcal S^H_e$ reproducing the same metric are related by one and only one local Lorentz transformation in $L_H$. Hence, the set of tetrads in $\S_e^H$ reproducing a given $g\in\S_g$ carries a free and transitive action of $L_H$, and the structure $\S_e^H\xrightarrow{\pi_g}\S_g$ is formally that of a principal $L_H$-bundle.

\subsubsection*{Residual symmetries}
Again, let us consider $e\in\S_e^H$ and ask what is the set $\mathfrak{a}_e^H$ of gauge symmetry generators inducing field-space transformations that are tangent to $\S_e^H$ at $e$. These must first of all be such that $\delta_{(\xi,\lambda)}c_g^i[g[e]]=0$, which constrains the vector-field part to be in $\mathfrak{g}_{g[e]}$. Then, they must be such that $\delta_{(\xi,\lambda)}c_e^j[e]=0$. As opposed to the case of $\S_e$ the Lorentz part $\lambda$ is then constrained, in a way that depends on the vector part $\xi$. In order to go further, let us assume that we can find a particular solution $\sigma_e[\xi]$ for $\lambda$. Then, the infinitesimal version of the conditions \eqref{eq:H-invariance} on the $c_e^j$ gives the important information that pure Lorentz generators $(0,\lambda)$ are also in $\mathfrak{a}_e^H$, if and only if $\lambda\in\mathfrak{l}_H\coloneqq\text{Lie}(L_H)$. As a result, general solutions to $\delta_{(\xi,\lambda)}c_e^j[e]=0$ are given in term of the particular one as $\lambda=\sigma[\xi]+\mu$ for $\mu\in\mathfrak{l}_H$. The set of tangent generators at $e$ is then finally
\be
\mathfrak{a}_e^H=\big\{(\xi,\sigma_e[\xi]+\mu)\,\vert\,\xi\in\mathfrak{g}_{g[e]},\,\mu\in\mathfrak{l}_H\big\}.
\ee
If we assume further that the resolution leading to the particular solution $\sigma_e[\xi]$ can be conducted uniformly in $\S_e^H$, then the set $\mathfrak{a}_e^H$ takes the same generic form in terms of $\sigma$ at every $e\in\S_e^H$. The residual symmetry algebra of $\S_e^H$ is then
\be
\mathfrak{a}^H\coloneqq\big\{(\xi,\lambda):\S_e^H\rightarrow\mathfrak{aut}(\P_{\vert U})\,\vert\,(\xi[e],\lambda[e])\in\mathfrak{a}_e^H\,,\forall e\in\S_e^H\big\}.
\ee

\subsubsection*{Projectable residual symmetries}
\label{sec:projectablegenerators}
As for $\S_e$, we now concentrate on projectable generators of $\mathfrak{a}^H$, whose vector part depends on $e$ only through $g[e]$. Using the same identification of $\xi\in\mathfrak{g}$ with $\xi\circ\pi_g$ as before and the particular solution $\sigma$, the set of projectable generators is
\be
\mathfrak{a}^H_\pi\coloneqq\big\{(\xi,\lambda)\in\mathfrak{a}^H\,\vert\,\xi\in\mathfrak{g}\big\}=\big\{(\xi,\sigma[\xi]+\mu)\,\vert\,\xi\in\mathfrak{g}\,,\mu\in\C^\infty(\S_e^H,\mathfrak{l}_H)\big\}.
\ee
As for $\mathfrak{a}_\pi$, a first observation is that for two elements $(\xi_1,\lambda_1)$ and $(\xi_2,\lambda_2)$, since $\delta_{(\xi_1,\lambda_1)}\xi_2=\delta_{\xi_1}\xi_2$ and similarly for $1\leftrightarrow2$, the vector field part of their bracket is simply $[\xi_1,\xi_2]_\star$, which also depends on $e$ only through $g[e]$. Hence $\mathfrak{a}^H_\pi$ is a subalgebra of $\mathfrak{a}_H$. The other similar result is that pure Lorentz generators of the form $(0,\mu)$ with $\mu\in\C^\infty(\S_e^H,\mathfrak{l}_H)$ form an ideal. Indeed, for any $(\xi,\lambda)\in\mathfrak{a}_\pi^H$ we have 
\be
[(\xi,\lambda),(0,\mu)]_\star=\big(0,\,\xi[\mu]+[\lambda,\mu]-\delta_{(\xi,\lambda)}\mu+\delta_{(0,\mu)}\lambda\big),
\ee
which does not have any vector field part thanks to projectability. At first glance it is not clear that the Lorentz part takes its values in $\C^\infty(\S_e^H,\mathfrak{l}_H)$. In particular, $H$ is not asked to be a normal subgroup so the bracket $[\lambda,\mu]$ is not ensured to be valued in $\mathfrak{h}\coloneqq\text{Lie}(H)$ pointwise. However, because the bracket lies in $\mathfrak{a}_\pi^H$ and has a vanishing vector field part, we are guaranteed by the property \eqref{eq:H-invariance} of the $c^j_e$ that that the sum of terms defining the Lorentz part is indeed in $\mathfrak{h}$ at each point. The last step to get a structure result similar to \eqref{eq:structureofapi} for $\mathfrak{a}_\pi^H$ is to see if $\mathfrak{g}$ can be realized as a subalgebra. This is where a generic obstruction arises.

\subsubsection*{Obstruction to the lift}
\label{sec:obstruction}
The realization of $\mathfrak{g}$ in $\mathfrak{a}_\pi$ above was obtained using the homomorphism $\xi\mapsto(\xi,0)$. This map is no longer available here, since generators of the form $(\xi,0)$ do not preserve the conditions $c^j_e[e]=0$ in general, so they are simply not in $\mathfrak{a}_\pi^H$. The natural candidate is then given by the particular solution $\sigma$ that we assumed to have found, namely the map $\xi\in\mathfrak{g}\mapsto\hat{\xi}\coloneqq(\xi,\sigma[\xi])\in\mathfrak{a}_\pi^H$. This time the map is well-defined, but it generically fails to be a Lie algebra homomorphism. Indeed, the bracket of the image of two elements $\xi_1$ and $\xi_2$ by this map has a vector field part $[\xi_1,\xi_2]_\star$, but nothing ensures that the Lorentz part is such that the whole bracket is the image of $[\xi_1,\xi_2]_\star$. In fact, we have
\be
[\hat{\xi}_1,\hat{\xi}_2]_\star=[(\xi_1,\sigma[\xi_1]),(\xi_2,\sigma[\xi_2])]_\star=\big([\xi_1,\xi_2]_\star,\,\sigma[[\xi_1,\xi_2]_\star]+\beta[\xi_1,\xi_2]\big)\neq\widehat{[\xi_1,\xi_2]_\star},
\ee
where $\beta[\xi_1,\xi_2]\in\C^\infty(\S_e^H,\mathfrak{l}_H)$ has no reason to vanish in general and prevents $\xi\mapsto\hat{\xi}$ from being a homomorphism. However, $\beta$ depends on the particular solution $\sigma$ we started from. Indeed, we argued that if $\sigma$ solves $\delta_{(\xi,\sigma[\xi])}c^j_e[e]=0$, then so does $\sigma'\coloneqq\sigma+\tau$ with $\tau[\xi]\in\C^\infty(\S_e^H,\mathfrak{l}_H)$. This defines a new map $\xi\mapsto\hat{\xi}'=(\xi,\sigma'[\xi])$, and a corresponding transformation $\beta\mapsto\beta'$ with $\beta'$ being defined as $\beta$, that is as the obstruction for $\hat\xi'$ to be a Lie algebra morphism. Writing $[X,Y]_\star^L$ for the Lorentz part of the modified bracket of $X$ and $Y$, and using $\tau[\xi]$ as a shorthand for $(0,\tau[\xi])$ when it appears in a bracket, the transformation for $\beta$ is explicitly
\be
\label{eq:betatransformation}
\beta[\xi_1,\xi_2]\mapsto\beta'[\xi_1,\xi_2]=\beta[\xi_1,\xi_2]+[\hat{\xi}_1,\tau[\xi_2]]_\star^L-[\hat{\xi}_2,\tau[\xi_1]]_\star^L+[\tau[\xi_1],\tau[\xi_2]]_\star^L-\tau[[\xi_1,\xi_2]_\star],
\ee
where the $\hat\xi_i$ are the ones associated with $\sigma$. 

Technically, $\beta$ is a non-abelian, field-dependent 2-cocycle \cite{alekseevsky2000extensions} and the formula above is its transformation under the change of particular solution $\sigma\mapsto\sigma+\tau$. The relevant question here is whether there exists such a transformation for which the cocycle is zero. In that case, $\hat\xi'$ would be a Lie algebra morphism and we would have succeeded in realizing the metric algebra $\mathfrak{g}$ as a subalgebra of the tetrad one $\mathfrak{a}^H_\pi$. This would require finding a certain $\tau$ for which \eqref{eq:betatransformation} vanishes, but it cannot be done without specifying the gauge conditions and the particular solution $\sigma$. In the next subsections, we follow a different route: instead of starting from a particular solution, we introduce a reference field and the associated comparison field and show that these directly provide a homomorphic lift of $\mathfrak g$ into $\mathfrak a^H_\pi$.
 
%%%%%%%%%%%%%%%%%%%%%%%%%%%%%%%%%%%%%%%%

\subsection{Reference and comparison fields}
\label{sec:referenceandcomparison}
The goal of this subsection is to introduce the tools that will replace the particular solution $\sigma$ of the tetrad gauge conditions used above.

\subsubsection*{Reference field}
Instead of looking for a particular solution for the Lorentz part of the generators, suppose that we can find, for every metric $g\in\S_g$, a tetrad $\underline{e}[g]$ which reproduces $g$ and satisfies the tetrad gauge conditions, and that this choice depends smoothly on $g$. We call such an assignment a \textit{reference field}. In the language of bundles, it is a smooth section of $\S_e^H\xrightarrow{\pi_g}\S_g$:
\be
\underline{e}:\S_g\to\S_e^H,\qquad g\mapsto\underline{e}[g],\qquad\pi_g\circ\underline{e}=\text{id}_{\S_g}.
\ee

\subsubsection*{Comparison field}
Note that in practice, a reference field is what one naturally constructs when solving the tetrad gauge conditions explicitly: a tetrad written as a functional of the metric only. But a generic tetrad $e\in\S_e^H$ might not be in the image of $\underline{e}$. In fact, because of the $L_H$-bundle structure of $\S_e^H$, we now that if $e$ produces a metric $g$, it is related to $\underline{e}[g]$ by one and only one $H$-valued local Lorentz transformation. More precisely, for every $e\in\S_e^H$, there exists a unique $\Lambda_0\in L_H$ such that 
\be
\label{eq:reference}
e=\Lambda_0[e]\,\underline{e}[\pi_g[e]].
\ee
The local Lorentz-valued field-space functional $\Lambda_0\in\C^\infty(\S_e^H,L_H)$ is called the \textit{comparison field}, and it is uniquely determined by the reference field $\underline{e}$. One can note the similarity between the notion of comparison field and that of dressing field in \textit{e.g.} \cite{franccois2021note}.

\subsubsection*{Transformation laws}
Let us now give the transformation laws for the two objects introduced above. First, note that the reference field $\underline{e}$ is a functional on the metric solution space $\S_g$. Differentiating $\pi_g\circ\underline{e}=\text{id}_{\S_g}$ gives that $\d\pi_g\circ\d\underline{e}=\text{id}_{T\S_g}$. Evaluating this identity on a metric variation $\delta_\xi g=\pounds_\xi g$ then gives $\d\pi_g\circ\d\underline{e}(\pounds_\xi g)=\d\pi_g(\delta_\xi\underline{e}[g])=\pounds_\xi g$. Next, as $\pi_g$ commutes with diffeomorphisms, we have that $\d\pi_g(\pounds_\xi \underline{e}[g])=\pounds_\xi\pi_g[\underline{e}[g]]=\pounds_\xi g$. These two results combine to give that $\d\pi_g(\delta_\xi\underline{e}-\pounds_\xi\underline{e})=0$, that is $\delta_\xi\underline{e}-\pounds_\xi\underline{e}\in\text{Ker}(\d\pi_g)$. The difference must then be a vertical vector of $\S_e$ at each point\footnote{And not one of $\S_e^H$, since $\delta_\xi\underline{e}$ is tangent to $\S_e^H$ but $\pounds_\xi\underline{e}$ is not, as it does not preserve the tetrad gauge conditions in general.}. This means that there exists a unique $\mathfrak{l}$-valued generator $\rho_{\underline{e}}$ such that for any field-dependent metric generator $\xi$,
\be
\label{eq:referencetransformation}
\delta_\xi\underline{e}=\pounds_\xi\underline{e}+\rho_{\underline{e}}[\xi]\,\underline{e}.
\ee
Note that in order to compute the Lorentz transformation $\rho_{\underline{e}}$ in practice, one only has to use the transformation of the metric under residual symmetries of $\S_g$ and compute $\delta_\xi\underline{e}$. In particular, no knowledge of the residual algebra of $\S_e^H$ is needed.

Next, what is the transformation of the comparison field $\Lambda_0$ under the action of a generic field-dependent generator $(\xi,\lambda)$? Writing it in matrix notation as $\Lambda_0=e\,\underline{e}^{-1}$, we can use that $\delta\Lambda_0=\delta e\,\underline{e}^{-1}-e\,\underline{e}^{-1}\,\delta\underline{e}\,\underline{e}^{-1}$. Using then the transformation laws \eqref{eq:transformationlaws} for $e$ and \eqref{eq:referencetransformation} for $\underline{e}$, we obtain
\be
\label{eq:comparisontransformation}
\delta_{(\xi,\lambda)}\Lambda_0=\lambda\,\Lambda_0-\Lambda_0\,\rho_{\underline{e}}[\xi]+\xi[\Lambda_0].
\ee

%%%%%%%%%%%%%%%%%%%%%%%%%%%%%%%%%%%%%%%%

\subsection{Structure theorem}
\label{sec:structuretheorem}
Here comes the main result of this section. Recall that we already obtained that $\mathfrak{a}_\pi^H$ contains an ideal, made from residual generators of the form $(0,\lambda)$, with vanishing vector field part. For a generic generator $(\xi,\lambda)$, we also argued that from the metric gauge conditions, its vector part should be in the residual metric algebra $\mathfrak{g}$. The problem appeared when imposing the preservation of the tetrad gauge conditions, namely when trying to solve $\delta_{(\xi,\lambda)}c^j_e=0$ for the Lorentz part $\lambda$, as a functional of $\xi$. In particular, we saw that even when assuming that we are able to find a particular solution $\sigma[\xi]$, the map $\xi\mapsto(\xi,\sigma[\xi])$ generically fails to be a Lie algebra homomorphism and therefore fails to realize $\mathfrak{g}$ as a subalgebra of $\mathfrak{a}_\pi^H$. 

Here, we assume that instead of a particular solution, we have a reference field $\underline{e}:\S_g\rightarrow\S_e^H$, and its associated comparison field $\Lambda_0:\S_e^H\rightarrow L_H$. A basic fact from the theory of principal bundles is that their local sections are in one-to-one correspondence with their local trivializations, see \textit{e.g.} \cite{Kobayashi1963Foundations}. In a sense, local sections can be used to coordinatize principal bundles. An instance of this is described in appendix \ref{app:geometry}, for the trivialization of the Lorentz bundle underlying tetrad formalism. Here, we face the same phenomenon: the section $\underline{e}$ defines uniquely $\Lambda_0$, which provides a map 
\be
\label{eq:trivialization}
\S_e^H\rightarrow\S_g\times L_H,\quad e\mapsto(\pi_g[e],\Lambda_0[e]),
\ee
with inverse $(g,\Lambda)\mapsto\Lambda\,\underline{e}[g]$. That is exactly a trivialization map for $\S_e^H$, associated with $\underline{e}$. It is then natural to ask which generators are horizontal with respect to the trivialization given by $\underline{e}$. That is, which are the projectable generators $(\xi,\lambda)$ with $\xi\in\mathfrak{g}$ such that $\delta_{(\xi,\lambda)}e=\delta_{(\xi,\lambda)}(\Lambda_0\,\underline{e})=\Lambda_0\,\delta_{(\xi,\lambda)}\underline{e}$, or equivalently, such that $\delta_{(\xi,\lambda)}\Lambda_0=0$. Using the transformation law \eqref{eq:comparisontransformation} for $\Lambda_0$, we get that $\lambda$ is uniquely defined by this condition as
\be
\label{eq:lift}
\lambda\coloneqq s[\xi]=\text{Ad}_{\Lambda_0}\rho_{\underline{e}}[\xi]-\xi[\Lambda_0]\Lambda_0^{-1}.
\ee
What remains to do is to show that this Lorentz transformation is precisely what we need in order to realize $\mathfrak{g}$ as a subalgebra of $\mathfrak{a}_\pi^H$, which then acquires the structure we expected.

\begin{proposition}
\label{prop:structureresult}
With all the notations introduced above, for any given $\xi\in\mathfrak{g}$, define $\hat\xi=(\xi,s[\xi])$. Then
\begin{enumerate}
\item $\hat\xi$ is an element of $\mathfrak{a}_\pi^H$;
\item the map $\xi\mapsto\hat\xi$ is an injective Lie algebra homomorphism from $\mathfrak{g}$ to $\mathfrak{a}_\pi^H$;
\item every element of $\mathfrak{a}_\pi^H$ can be written as $\hat\xi+(0,\mu)$, with $\xi\in\mathfrak{g}$ and $\mu\in\C^\infty(\S_e^H,\mathfrak{l}_H)$, and we have
\be
[\hat\xi,(0,\mu)]_\star=(0,-\delta_{\hat\xi}\mu).
\ee
\end{enumerate}
Consequently, the algebra of projectable residual symmetries $\mathfrak{a}_\pi^H$ has a semi-direct sum structure
\be
\mathfrak{a}_\pi^H=\mathfrak{g}\rsemidirsum\C^\infty(\S^H_e,\mathfrak{l}_H),
\ee
where $\mathfrak{g}$ is identified with its image under the lift and acts semi-directly on the ideal by $\mu\mapsto-\delta_{\hat\xi}\mu$.
\end{proposition}

\begin{proof}
(1) First, $\hat\xi$ is projectable by definition. Then, for any given $e\in\S^H_e$, consider the map $\Phi_e:g'\in\S_g\mapsto\Lambda_0[e]\,\underline{e}[g']\in\S^H_e$, which takes values in $\S^H_e$ since $\underline{e}$ does and since $\Lambda_0[e]\in L_H$. Writing $g=\pi_g[e]$, we have $\Phi_e(g)=e$ and then since $\delta_{\hat\xi}\Lambda_0=0$, we have $\delta_{\hat\xi}e=\Lambda_0\,\delta_{\xi}\underline{e}=\d\Phi_e(\pounds_\xi g)$. This is the pushforward of a tangent vector to $\S_g$ at $g$, hence a tangent vector to $\S^H_e$ at $e$. This means that $\hat\xi\in\mathfrak{a}_\pi^H$.

(2) Injectivity is clear from the definition of $\hat{\xi}$. Then, given two lifted vectors $\hat{\xi_1}$ and $\hat{\xi_2}$, we have that $[\delta_{\hat{\xi_1}},\delta_{\hat{\xi_2}}]\,\Lambda_0=0$, so that by the representation theorem \eqref{eq:representationtheorem}, $\delta_{[\hat{\xi_1},\hat{\xi}_2]_\star}\,\Lambda_0=0$. As the vector part of their bracket is $[\xi_1,\xi_2]_\star$ and by uniqueness of the solution to $\delta\Lambda_0=0$, the Lorentz part must exactly be $s[[\xi_1,\xi_2]_\star]$. In other words, we have $[\hat{\xi_1},\hat{\xi}_2]_\star=\widehat{[\xi_1,\xi_2]_\star}$.

(3) First, for $X\in\mathfrak{a}_\pi^H$ with vector part $\xi$, the generator $X-\hat\xi$ preserves the tetrad gauge conditions
and has vanishing vector part, so that it is of the form $(0,\mu)$ with $\mu$ valued in $\mathfrak{l}_H$. The decomposition $X=\hat\xi+(0,\mu)$ is clearly unique. Next, we can compute, using $\delta_{(0,\mu)}\Lambda_0=\mu\,\Lambda_0$ and $\delta_{(0,\mu)}\rho_{\underline{e}}[\xi]=0$ that
\be
\notag
\delta_{(0,\mu)}s[\xi]&=\delta_{(0,\mu)}\big{(}\Lambda_0\,\rho_{\underline{e}}[\xi]\,\Lambda_0^{-1}-\xi[\Lambda_0]\,\Lambda_0^{-1}\big{)}\\\notag
&=\mu\,\Lambda_0\,\rho_{\underline{e}}[\xi]\,\Lambda_0^{-1}-\Lambda_0\,\rho_{\underline{e}}[\xi]\,\Lambda_0^{-1}\mu-\xi[\mu\,\Lambda_0]\,\Lambda_0^{-1}+\xi[\Lambda_0]\,\Lambda_0^{-1}\,\mu\\\notag
&=[\mu,\text{Ad}_{\Lambda_0}\,\rho_{\underline{e}}[\xi]]-\xi[\mu]-[\mu,\xi[\Lambda_0]\,\Lambda_0^{-1}]\\\notag
&=[\mu,s[\xi]]-\xi[\mu],
\ee
from which we deduce the bracket $[\hat{\xi},(0,\mu)]_\star=(0,\xi[\mu]+[s[\xi],\mu]-\delta_{\hat{\xi}}\mu+\delta_{(0,\mu)}s[\xi])=(0,-\delta_{\hat{\xi}}\mu)$.
\end{proof}

Before summarizing this section and examining its upshots, let us make some remarks, following directly from the proposition above.
\begin{itemize}
\item In the language of the previous subsection, what we just showed is that any given particular solution $\sigma$ to the preservation of the tetrad gauge conditions can be conveniently redefined by $\tau=s-\sigma$ in order to remove its associated cocycle.

\item If the elements of the ideal $\C^\infty(\S^H_e,\mathfrak{l}_H)$ of pure $\mathfrak{l}_H$-valued Lorentz transformations are taken to be field-independent, then the semi-direct action becomes trivial and the residual symmetry algebra becomes a direct sum. In particular, suppose that we have a parameterization $\xi[g,\alpha]$, $\alpha\in P$ of the metric residual symmetry algebra, as defined in section \ref{sec:solutionspaces}, with a Lie algebra structure on $P$ denoted by $\mathfrak{g}^\text{par}$. Then, we get for free a parameterization of $\mathfrak{a}_\pi^H$, with the field-independent pure residual Lorentz transformations as parameters for the ideal part. The full residual symmetry algebra of parameters is then the direct sum $\mathfrak{g}^\text{par}\oplus\mathfrak{l}_H$. What this means is that in that setting, the existence of such a parameterization is a property of the metric solution space only, and the passage to tetrad variables introduces no new obstruction.

\item The lift $\hat\xi$ reduces to the Kosmann lift described in appendix \ref{app:kos} for metric isometry generators. Indeed, suppose that $\xi\in\mathfrak{g}$ happens to be such that $\xi[g]$ is a Killing vector of $g$ so that $\pounds_{\xi[g]}g=0$. Then, if $e\in\S_e^H$ projects onto $g$, we have that $\delta_{\hat{\xi}}e=\Lambda_0[e]\,\delta_{\xi}\underline{e}=\Lambda_0[e] \,\d \underline{e}(\pounds_{\xi}g)=0$. Finally, since for a given vector part $\xi$, there is at most one Lorentz part such that $\delta_{(\xi,\lambda)}e=0$, which exists if and only if $\xi$ is a Killing vector of $g[e]$ and is then given by the Kosmann lift $\K^e_\xi$, the Lorentz part of $\hat\xi$ coincides at $e$ with that of $\K^e_\xi$, independently of the choice of reference field.

\item Finally, note that the expression of the lift $\xi\mapsto\hat\xi$ depends on the reference field. Indeed, for any $h\in\C^\infty(\S_g,L_H)$, it is clear that $\underline{e}'\coloneqq h\,\underline{e}$ is another valid reference field. The associated comparison field is then given by $\Lambda_0'=\Lambda_0 h^{-1}$. Computing the variation of the new reference field, a few lines give $\delta_\xi\underline{e}'=\pounds_\xi\underline{e}'+\rho_{\underline{e}'}[\xi]\,\underline{e}'$, with the new associated Lorentz transformation
\be
\rho_{\underline{e}'}[\xi]=\text{Ad}_h\rho_{\underline{e}}[\xi]+(\delta_\xi h-\pounds_\xi h)h^{-1}.
\ee
Plugging this expression into \eqref{eq:lift}, we find that the lift associated with the new reference field tetrad $\underline{e}'$ is given by $\xi\mapsto\hat{\xi}'=(\xi,s'[\xi])$ with
\be
s'[\xi]=s[\xi]+\text{Ad}_{\Lambda_0}(h^{-1}\delta_\xi h).
\ee
Both lifts are injective Lie algebra homomorphisms, but they define different, though isomorphic, realizations of $\mathfrak{g}$ in $\mathfrak{a}_\pi^H$. Note that if $h$ is field-independent, then the lift does not change.
\end{itemize}

\subsection{Summary and upshots}
\label{sec:summ}
Let us summarize the results of this section. Starting from a metric solution space $\S_g$ with residual symmetry algebra $\mathfrak{g}$, we have described it in tetrad variables, first without any constraint on the Lorentz frame, then with additional tetrad gauge conditions leaving a residual subgroup $H$ of local Lorentz transformations. In both cases, the tetrad solution space is a principal bundle over $\S_g$, and the projectable residual symmetries form a Lie algebra $\mathfrak{a}_\pi^H$, containing the $\mathfrak{l}_H$-valued field-dependent residual Lorentz generators as an ideal. Without tetrad conditions, $\mathfrak{g}$ is trivially realized as a subalgebra of $\mathfrak{a}_\pi^H$ through $\xi\mapsto(\xi,0)$. With tetrad conditions, the compensating Lorentz transformation associated with metric generators, obtained as a particular solution to the preservation of the tetrad gauge conditions, generically spoils this realization through a field-dependent cocycle $\beta$. We have shown that a reference field, \textit{i.e.} a section of the bundle $S_e^H\xrightarrow{\pi_g}\S_g$, always provides a lift free of this cocycle, so that $\mathfrak{a}_\pi^H=\mathfrak{g}\rsemidirsum\C^\infty(\S^H_e,\mathfrak{l}_H)$, with a direct sum for field-independent generators in the ideal. Let us now make some comments on this result.

\paragraph{Generality.} This result does not depend on the specific form of the metric solution space, on the metric and tetrad gauge and boundary conditions, nor on the residual subgroup $H$, which does not need to be abelian or normal. It only relies on the covariance of the dynamics, on the principal bundle
structure of $\S_e$ over $\S_g$, and on the existence of a reference field. In particular, it applies to any Lorentz-invariant and diffeomorphism-covariant Lagrangian, including the Holst, Euler and Pontryagin terms which can be added to the action since they do not modify the solution space.

\paragraph{Practical upshot.} The construction provides a shortcut to compute the residual symmetries of a tetrad solution space in a form adapted to the metric ones. One only needs the residual symmetries of the metric solution space and the transformation laws of the metric, together with a reference tetrad $\underline{e}[g]$, which is what one naturally writes down when solving the tetrad gauge conditions. The transformation of $\underline{e}$ then gives $\rho_{\underline{e}}$, and the lift follows from \eqref{eq:lift}. In particular, it is not necessary to solve the preservation of the tetrad gauge conditions for the Lorentz part of tetrad generators, nor to compute the cocycle of a naive lift. This will be illustrated in the next section.

\paragraph{Robustness of metric asymptotic symmetries.} From a physical point of view, the result states that the residual symmetry algebra of a metric solution space survives the passage to a first-order description provided that we are able to find a reference tetrad, which can in principle always be done at least locally in the metric solution space. Also, if the metric description admits a parameterization, so does the tetrad one, and the passage to tetrads introduces no new obstruction.

\paragraph{Non-projectable generators.} We have restricted to projectable generators. As explained above, residual generators whose vector part depends on the tetrad beyond the metric exist, but they do not induce well-defined transformations of the metric solution space, and therefore do not realize metric residual symmetries. They are not relevant for the question addressed here.
 
\paragraph{Upshots for charges.} Although charges associated with residual symmetries (in the sense of \cite{Iyer:1994ys, Wald:1999wa}) are beyond the scope of this paper, the structure result has direct consequences for them. Let us assume that we are in a setting where residual symmetries are represented by their charges through a bracket of the form
\be
\{Q_{X_1},Q_{X_2}\}=Q_{[X_1,X_2]_\star}+K_{X_1,X_2},
\ee
where $K$ is a possible field-dependent 2-cocycle, as is the case provided that we have a convenient integrable part and flux split, see \textit{e.g.}  \cite{Wald:1999wa, Barnich:2001jy, Barnich:2007bf, Rignon-Bret:2024wlu}. Then:
\begin{itemize}
\item With a lift associated with a particular solution $\sigma$, the bracket of the charges of two naively lifted metric generators reads $\{Q_{(\xi_1,\sigma[\xi_1])},Q_{(\xi_2,\sigma[\xi_2])}\}=Q_{([\xi_1,\xi_2]_\star,\sigma[[\xi_1,\xi_2]_\star])}+Q_{(0,\beta[\xi_1,\xi_2])}+K$: the charges of metric residual symmetries do not close among themselves, and residual Lorentz charges appear in their algebra through $\beta$. With the lift associated with a reference field, this term is absent, and the charges of lifted metric generators close among themselves, up to $K$.
\item For field-independent $\mu\in\mathfrak{l}_H$, $[\hat\xi,(0,\mu)]_\star=0$, so that $\{Q_{\hat\xi},Q_{(0,\mu)}\}=K_{\hat\xi,(0,\mu)}$: the charges of lifted metric generators and of field-independent residual Lorentz transformations commute, up to $K$.
\item Under a field-dependent change of reference field, the charges of lifted generators change as $Q_{\hat\xi'}=Q_{\hat\xi}\,+\,Q_{(0,\text{Ad}_{\Lambda_0}(h^{-1}\delta_\xi h))}$. The charges of metric symmetries computed in a first-order formulation therefore depend on the choice of reference frame whenever the residual Lorentz charges are non-trivial.
\end{itemize}
These effects have to be taken into account whenever the residual Lorentz transformations have non-vanishing charges. Whether it is the case or not depends on the residual group $H$, the boundary conditions and on the Lagrangians entering the symplectic structure, see \textit{e.g.} \cite{DePaoli:2018erh, Godazgar:2020gqd, Freidel:2020xyx, Godazgar:2022foc}.

%%%%%%%%%%%%%%%%%%%%%%%%%%%%%%%%%%%%%%%%
%%%%%%%%%%%%%%%%%%%%%%%%%%%%%%%%%%%%%%%%

\section{BMSW-$u$ solution space with residual U(1) Lorentz symmetry}
\label{sec:BMSW-u-U(1)}

%%%%%%%%%%%%%%%%%%%%%%%%%%%%%%%%%%%%%%%%

\subsection{Metric solution space}
\label{sec:metricsolutionspace}
The local metric solution space $\S_g$ that we consider in this part is that of \cite{Compere:2018ylh, Freidel:2021fxf, Geiller:2024bgf}, but with arbitrary time dependance in the leading angular metric defined just below. It is written in local coordinates $x^\mu=(u,r,x^a)$. Every metric in $\S_g$ has a line element of the Bondi form \cite{Bondi:1962px, Sachs:1962wk} (see \textit{e.g.} \cite{Madler:2016xju} for a review)
\be
\d s^2=\f{V}{r}e^{2B}\d u^2-2e^{2B}\d u\,\d r+\gamma_{ab}(\d x^a-U^a\d u)(\d x^b-U^b\d u),
\ee
where we assume that the functions $B(u,r,x^a)$, $V(u,r,x^a)$, $U^a(u,r,x^b)$, as well as the transverse two-dimensional metric $\gamma_{ab}(u,r,x^c)$, admit an expansion in inverse powers of $r$ as it goes to infinity, which is determined via Einstein equations. Such a metric is asked to satisfy the Bondi-Sachs gauge conditions
\be
\label{eq:metricgauge}
g_{rr}=0,
\q\q
g_{ra}=0,
\q\q
\partial_r\left(\f{\det\gamma}{r^4}\right)=0.
\ee
While the first two can be read off from of the line element and are algebraic in the metric components, we add the third one, which is differential. It is automatically solved if we express the expansion of the transverse metric $\gamma_{ab}$ as
\be
\label{eq:gammaexpansion}
\gamma_{ab}=r^2q_{ab}\sqrt{1+\f{[\C\C]}{2r^2}}+r\C_{ab},
\q\q
r\C_{ab}\coloneqq rC_{ab}+\sum_{n=1}^\infty\f{E^n_{ab}}{r^n},
\ee
where $q_{ab}$ is the leading angular metric and the so-called shear tensor $C_{ab}$ as well as every $E^n_{ab}$ are all trace-free in $q_{ab}$, and using the notation $[\C\C]=\C^{ab}\C_{ab}$, the indices being raised and lowered by $q_{ab}$. Note that the latter is arbitrary, and in particular $\partial_uq_{ab}\neq0$. The determinant condition implies $\sqrt{\gamma}=r^2\sqrt{q}$, and therefore $\sqrt{-g}=r^2e^{2B}\sqrt{q}$. We also have $\D_aV^a=D_aV^a$ for any vector $V^a$, where $\D_a$ and $D_a$ are the derivatives with respect to $\gamma_{ab}$ and $q_{ab}$ respectively. Note also that setting the term of order zero in the expansion of $\C_{ab}$ to zero is exactly what removes the logarithmic branch \cite{Chrusciel:1993hx, Geiller:2022vto} in the expansion when performing the resolution whose results are presented below.

To the gauge conditions \eqref{eq:metricgauge}, we add the following supplementary boundary conditions, constraining the fall-offs in the radial expansion of the non-vanishing metric coefficients:
\be
\label{eq:boundaryconditions}
g_{uu}=\O(r),
\q\q
g_{ur}=-1+\O(r^{-2}),
\q\q
g_{ua}=\O(1),
\q\q
g_{ab}=\O(r^2).
\ee
Note that the last one is coherent with the expansion of \eqref{eq:gammaexpansion}, the latter being the only one compatible both with this fall-off and with the determinant condition.

It now remains to solve the vacuum Einstein equations $G_{\mu\nu}=0$, within the chosen gauge and while preserving the boundary conditions. First, the four hypersurface equations $G_{r\mu}=0$ give the expansion of the parameters $B$, $V$ and $U^a$. These equations are solved at the following orders:
\be
G_{rr}=\O(r^{-7}),
\q\q
G_{ra}=\O(r^{-5}),
\q\q
G_{ru}=\O(r^{-5}),
\ee
by imposing the following expansion:
\bsub
\be
B&=\f{B_2}{r^2}+\f{B_3}{r^3}+\f{B_4}{r^4}+\O(r^{-5}),\\
U^a&=\f{U^a_2}{r^2}+\f{U^a_3}{r^3}+\f{U^a_4}{r^4}+\O(r^{-5}),\\
V&=V_{+2}r^2+V_{+1}r+V_0+\f{V_1}{r}+\O(r^{-2}),
\ee
\esub
where
\bsub
\be
B_2&=-\f{1}{32}[CC],\\
B_3&=0,\\
B_4&=-\f{3}{32}[CE^1]+\f{1}{128}[CC]^2,\\
U^a_2&=-\f{1}{2}D_bC^{ab},\\
U^a_3&=N^a,\\
U^a_4&=\f{3}{4}D_bE_1^{ab}-\f{3}{4}C^{ab}N_b-\f{1}{16}C^{ab}\partial_b[CC]+\f{1}{64}[CC]D_bC^{ab},\\
V_{+2}&=-\partial_u\ln\sqrt{q},\\
V_{+1}&=-\f{R}{2},\\
V_0&=2M,\\
V_1&=\f{1}{2}D_aN^a+\f{1}{8}D_aC_{bc}D^aC^{bc}-\f{1}{2}D_aC^{ab}D^cC_{cb}+\f{3}{32}\big(R-D^2\big)[CC].
\ee
\esub
Here $R$ is the Ricci scalar of $q_{ab}$ and $D^2=D^aD_a$ is the Laplacian. The radial integration constants are the angular momentum and mass aspects appearing in $U^a_3=N^a(u,x^b)$ and $V_0=2M(u,x^a)$.
The remaining Einstein equations contain time evolution constraints to impose on some of the data. We can start by solving the trace-free part of the angular Einstein equations, defined as $G_{ab}^\tf=G_{ab}-\f{1}{2}\gamma_{ab}G$, with $G=\gamma^{ab}G_{ab}$. At order $\O(r)$, this equation imposes that\footnote{Here and after, we denote the symmetric and trace-free part of a tensor $T_{ab}$ by $T_{\la ab\ra}=T_{(ab)}-\f{1}{2}q_{ab}T_{cd}q^{cd}$.}
\be
\partial_uq_{ab}-q_{ab}\partial_u\ln\sqrt{q}=(\partial_uq)_{\la ab\ra}\stackrel{!}{=}0.
\ee
This means that the time dependence of $q_{ab}$ must be in a conformal factor, or in other words that
\be
q_{ab}(u,x^c)=e^{\psi(u,x^a)}Q_{ab}(x^c).
\ee
The next two orders of $G_{ab}^\tf$ are then automatically solved and, starting at order $\O(r^{-2})$, the equation contains an infinite tower of evolution equations for the angular aspects $E_{ab}^n$ entering the subleading expansion of the angular metric. It can then be shown \cite{Geiller:2022vto} that when the hypersurface equations and the trace-free angular equation are solved, then the trace part also vanishes, so that $G_{ab}=0$ is achieved. In fact, at this stage, the only remaining information in the Einstein equations are contained in the order $\O(r^{-2})$ of $G_{uu}$ and $G_{ua}$, which encode respectively the evolution equations for the mass and angular momentum aspects. They can be found in appendix \ref{app:evolutionequation}. 

At the end of the day, all this provides a parameterization of the metrics in the solution space $\S_g$ in terms of two types of data. The unconstrained data are $\big(\psi, Q_{ab}, C_{ab}\big)$, that is the conformal factor containing the time evolution of the leading angular metric, its time-independent part, and the shear tensor. The constrained data are $\big(M, N_a, E_{ab}^n\big)$, that is the mass and angular momentum aspects as well as the subleading angular aspects, which all have their own evolution equation.

%%%%%%%%%%%%%%%%%%%%%%%%%%%%%%%%%%%%%%%%

\subsection{Tetrad solution space}
We now want to obtain a tetrad description of the metric solution space of the previous subsection. For each metric in $\S_g$, we have the whole local Lorentz freedom to describe it in terms of a tetrad. Fixing part of this freedom is first motivated by the fact that writing a tetrad solution space concretely while keeping all of it quickly becomes cumbersome. Then, obtaining such an explicit writing allows for a precise study of the residual Lorentz freedom selected by the partial gauge fixing. Here, we will allow for and study a residual U(1) Lorentz symmetry. 

More precisely, the Lorentz gauge fixing that we will be using starts like the one in \textit{e.g.} \cite{Freidel:2021fxf, Godazgar:2022foc}. Using first Newman--Penrose notations, we are searching for a tetrad $e_{\text{NP}}^I=(\ell,n,m,\bar{m})$, with internal metric $\eta^{\text{NP}}$ such that $\eta^{\text{NP}}_{01}=-1$ and $\eta^{\text{NP}}_{23}=1$. What we impose is then that $\ell=\partial_r$, and furthermore that $m$ is tangent to the angular two spheres, so that $m=m^a\partial_a$. Asking then this tetrad to produce a given $g_{\mu\nu}\in\S_g$, that is that $g^{\mu\nu}=-2\ell^{(\mu}n^{\nu)}+2m^{(\mu}\bar{m}^{\nu)}$, imposes
\be
\ell=\partial_r,
\quad\;\
n= e^{-2B}\left(\partial_u+\f{V}{2r}\partial_r+U^a\partial_a\right),
\quad\;\
\gamma^{ab}=2m^{(a}\bar{m}^{b)}.
\ee
We see that $\ell$ and $n$ are fixed, but $m$ is only required to be a dyad for $\gamma_{ab}$. In terms of Newman-Penrose formalism (see \textit{e.g.} \cite{Chandrasekhar:1985kt}), this freezes the class I, II and the boost part of the class III Lorentz transformations. See also \textit{e.g.} \cite{Barnich:2019vzx, Geiller:2024bgf} for related discussions about this gauge fixing. The condition on $m$ of being a dyad for $\gamma_{ab}$ is invariant under the U(1) rotation part of class III Lorentz transformations\footnote{In fact, this condition is invariant under $O(2)$, but respecting a given orientation excludes the reflection $m\leftrightarrow\bar m$ and leaves only a U(1) invariance.}, and this is precisely the freedom we want to leave explicit. In fact, if $\underline{m}$ is a given dyad for $\gamma_{ab}$, any admissible $m$ will be written as $m=e^{i\varphi}\underline{m}$, for a unique U(1) angle $\varphi$. We know such an explicit reference dyad $\underline{m}$, and we will use it to write $m$ as follows:
\be
\label{eq:referencem}
m\coloneqq e^{i\varphi}\underline{m},
\quad\;\
\underline{m}\coloneqq\sqrt{\f{\gamma_{\theta\theta}}{2\gamma}}\left(\f{\sqrt{\gamma}+i\gamma_{\theta\phi}}{\gamma_{\theta\theta}}\,\partial_\theta-i\partial_\phi\right).
\ee

In order to stick to notations of the literature \cite{Freidel:2021fxf, Godazgar:2022foc} and check several results against it, we will in fact switch to the use of a real dyad $E_i=(E_2,E_3)$, simply defined by
\be\label{real frame}
E_2\coloneqq\f{1}{\sqrt{2}}(\bar{m}+m),
\q\q
E_3\coloneqq\f{1}{i\sqrt{2}}(\bar{m}-m),
\ee
and such that $\gamma^{ab}=E^a_iE^b_j\delta^{ij}$.
We can then write a new version of \eqref{eq:referencem} as
\be\label{E with phase}
E_i=
\begin{pmatrix}
\cos\varphi&\sin\varphi\\
-\sin\varphi&\cos\varphi
\end{pmatrix}{\raisebox{-0.6em}{\!\!{$_{ij}$}}}\,
\underline{E}_j,
\q\q
\underline{E}_2=\f{1}{\sqrt{\gamma_{\theta\theta}}}\partial_\theta,
\q\q
\underline{E}_3=-\sqrt{\f{\gamma_{\theta\theta}}{\gamma}}\left(\f{\gamma_{\theta\phi}}{\gamma_{\theta\theta}}\,\partial_\theta-\partial_\phi\right),
\ee
where $\underline{E}$ is the real dyad constructed from $\underline{m}$. The complete real tetrad is denoted $e^I=(\ell,n,E_2,E_3)$ and is such that $g^{\mu\nu}=e^\mu_I e^\nu_J\eta^{IJ}$ with the internal metric such that $\eta_{01}=-1$ and $\eta_{22}=\eta_{33}=1$. 

Let us summarize the tetrad gauge and boundary conditions. First, we started with the tetrad conditions that $\ell=\partial_r$, and that $m$ (equivalently, $E_i$) is tangent to the angular two spheres. A convenient and equivalent set of conditions can be written in terms of the tetrad with downstair spacetime indices as
\be
\label{eq:tetradgauge}
e^1_r=-1,
\q\q
e^1_a=0,
\q\q
e^0_a=0.
\ee
Then, as boundary conditions, we impose that the angular dyad $E_i$ admits an expansion in inverse powers of $r$ and further that $E_i=\O(r^{-1})$. This condition can be written as $e^i_a=\O(r^2)$, or in terms of the new solution space parameter $\varphi$, as
\be
\label{eq:anglefalloff}
\varphi=\O(1),
\ee
which discards residual U(1) rotations that blow up at infinity. The set of tetrads reproducing the metric of $\S_g$ and satisfying these gauge and boundary conditions is denoted by $\S_e^{\text{U(1)}}$.

As we want to work on-shell of the equations of motion for the Cartan--Palatini Lagrangian, the spin connection is determined by the tetrad through $\omega_\mu^{IJ}[e]=e^{\alpha I}\nabla_\mu e^J_\alpha$. The explicit expressions of its coefficients are written in appendix \ref{app:spinconnection}.

At this stage, we have succeeded in finding, for any metric $g_{\mu\nu}\in\S_g$, a set of tetrads $e^I[g,\varphi]$ reproducing it, all related by a unique local U(1) transformation. Note that $\varphi$ appears as the parameter for this residual freedom, and that we have it at our disposal thanks to the introduction of the reference dyad $\underline{m}$. In fact, if we define $\underline{e}^I[g] \coloneqq e^I[g,\varphi=0]=(\ell,n,\underline{E}_2,\underline{E}_3)$ and if $\Lambda_0[\varphi]$ is the Lorentz transformation corresponding to the rotation of the dyad by an angle $\varphi$, then $e^I[g,\varphi]=\Lambda_0[\varphi]\,\underline{e}^I[g]$ is the specification of equation \eqref{eq:reference} to the current case. In fact, in the language of section \ref{sec:3}, we are already working in the trivialization of the principal bundle $\S^{\text{U}(1)}_e\xrightarrow{\pi_g}\S_g$, given by the reference tetrad $\underline{e}^I$, as defined in equation \eqref{eq:trivialization}.  Finally, note that the origin of the angles entering the expansion of $\varphi$ can be shifted in order to define another valid reference tetrad. Another choice of reference is discussed in appendix \ref{app:alternativedyad}.

%%%%%%%%%%%%%%%%%%%%%%%%%%%%%%%%%%%%%%%%

\subsection{Metric asymptotic symmetries}
Let us present here the asymptotic\footnote{From now on, we switch from ``residual'' to ``asymptotic'', as is customary in the case where the boundary conditions constrain the asymptotic structure of the spacetimes under consideration, here at null infinity.} symmetries of the metric solution space $\S_g$. They are generated by vector fields $\xi$ which first preserve the gauge conditions
\be
g_{rr}=0,
\q\q
g_{ra}=0,
\q\q
\partial_r\left(\f{\det\gamma}{r^4}\right)=0.
\ee
This imposes the following form:
\be
\xi^u=f,
\q\q
\xi^r=\f{r}{2}\big(U^a\partial_af-2W-D_aI^a\big),
\q\q
\xi^a=Y^a+I^a=Y^a-\int_r^\infty e^{2B}\gamma^{ab}\partial_bf\de r',
\ee
where $f(u,x^a)$, $W(u,x^a)$, $Y^a(u,x^b)$ are for the moment unconstrained. Preserving the boundary condition $g_{ur}=-1+\O(r^{-2})$ then imposes that $W=\partial_uf$, while preserving $g_{ua}=\O(1)$ imposes that $\partial_uY=0$. The two remaining boundary conditions $g_{uu}=\O(r)$ and $g_{ab}=\O(r^2)$ are then automatically preserved. Note that, as we allow for an arbitrary time-dependence in the (conformal factor of) the leading angular metric, we do not have to impose $\partial_uW=0$ as in \textit{e.g.} \cite{Freidel:2021fxf}. We see that these asymptotic symmetry vectors exhibit field-dependence and that they admit a parameterization, in the sense of section \ref{sec:solutionspaces}, by pairs $\big(f(u,x^a), Y^a(x^b)\big)$. 

By acting on the metric with the Lie derivative, one can find the transformation laws for the metric parameterizing functions $B$, $V$, $U^a$ and $\gamma_{ab}$. These are
\bsub
\label{eq:expansionfreetransformation}
\be
\delta_\xi B&=\big(f\partial_u+\xi^r\partial_r+\pounds_{(I+Y)}\big)B+\f{1}{2}\big((\partial_u+U^a\partial_a)f+\partial_r\xi^r\big),\\
\delta_\xi V&=\big(f\partial_u+\xi^r\partial_r+\pounds_{(I+Y)}\big)V-\big(\f{V}{r}+\partial_r+2r(\partial_u+U^a\partial_a)\big)\xi^r+V(\partial_u+U^a\partial_a)f,\\
\delta_\xi U^a&=\big(f\partial_u+\xi^r\partial_r+\pounds_{(I+Y)}\big)U^a+\big(U^aU^b\partial_b-\f{V}{r}e^{2B}\partial^a\big)f+e^{2B}\partial^a\xi^r-\partial_u\xi^a,\\
\delta_\xi \gamma_{ab}&=\big(f\partial_u+\xi^r\partial_r+\pounds_{(I+Y)}\big)\gamma_{ab}-2U_{(a}\partial_{b)}f\label{eq:gammatransfo}.
\ee
\esub
By looking at the radial expansion of these transformation laws, one can find that of all the actual metric solution space parameters. Some of these are written in appendix \ref{app:transformationlaws}. As we can work mainly without any expansion here, the formulas above are sufficient. 

With these transformation laws and the modified bracket \eqref{eq:vectmodifiedbracket}, we are ready to compute the algebra for the residual symmetry generators $\xi(f,Y)$ of $\S_g$. In fact, as we have a parameterization, we can consider field-independent parameters and see if the modified bracket makes them close among themselves in a field-independent way. It is the case, as we find the algebra for field-independent parameters to be given by $[\xi(f_1,Y_1),\xi(f_2,Y_2)]_\star\coloneqq\xi(f_{12},Y_{12})$, with
\be
Y_{12}&=[Y_1,Y_2],\\
f_{12}&=Y_1[f_2]-Y_2[f_1]+f_1\partial_uf_2-f_2\partial_uf_1.
\ee
This is the $\mathfrak{bmsw}_4$ algebra for arbitrary time-dependence in the leading angular metric, hence the name BMSW-$u$ for the solution space itself. The next question is that of the embedding of this algebra in that of the tetrad asymptotic symmetries.

%%%%%%%%%%%%%%%%%%%%%%%%%%%%%%%%%%%%%%%%

\subsection{Tetrad asymptotic symmetries}
\subsubsection{Naive lift and cocycle obstruction}
Let us now discuss the residual symmetries of the tetrad solution space $\S_e^{\text{U(1)}}$. Let us first recall that the symmetry generators in tetrad formalism take the form of pairs $(\xi,\lambda)$ of diffeomorphism and local Lorentz generators. Under the action of such a couple, a tetrad transforms as 
\be\label{full transformation of tetrad}
\delta_{(\xi,\lambda)}e^I_\mu=\pounds_\xi e^I_\mu+{\lambda^I}_Je^J_\mu.
\ee

Now, given that any tetrad in $\S_e^{\text{U(1)}}$ is designed to produce a metric in $\S_g$, it is subject to the metric gauge \eqref{eq:metricgauge} and boundary conditions \eqref{eq:boundaryconditions}, expressed in tetrad terms by writing explicitly the metric as a function of the tetrad. We have that $\delta_{(\xi,\lambda)}g_{\mu\nu}[e]=\pounds_\xi g_{\mu\nu}[e]$, for any Lorentz generator $\lambda$, so that these metric conditions only constrain the vector field part $\xi$ of admissible residual pairs of generators to be a symmetry generator of $\S_g$, that is to be of the form $\xi(f,Y)$ described in the previous paragraph. Note that at the level of metric parameters, we will never consider them as field-dependent. In particular, the restriction to projectable tetrad generators as defined in section \ref{sec:projectablegenerators} is automatic.

In order to find the right Lorentz part and since we have a reference tetrad at our disposal we could use directly the results of section \ref{sec:structuretheorem}. We choose here to first present the pedestrian way, consisting in finding a particular solution to the preservation of the tetrad gauge conditions, in order to illustrate the problem we encounter with this naive method at the level of the algebra. Let us then first constrain the Lorentz generator by the supplementary tetrad gauge conditions \eqref{eq:tetradgauge}, namely
\be
e^1_r=-1,
\q\q
e^1_a=0,
\q\q
e^0_a=0.
\ee
Asking the preservation of these gauge conditions by the joint action of a given metric residual symmetry generator $\xi$ and a Lorentz generators $\lambda$, that is asking that the components involved do not vary through \eqref{full transformation of tetrad}, determines uniquely part of $\lambda$ as a function of $\xi$:
\be
\label{eq:lambdas}
\lambda^{01}=-\partial_r\xi^r,
\q\q
\lambda^{0i}=e^{2B}E^a_i\partial_af,
\q\q
\lambda^{1i}=E^a_i\left(\partial_a\xi^r-\f{V}{2r}\partial_af\right).
\ee
With this, we recover the formulas of \cite{Godazgar:2020kqd, Godazgar:2022foc, Freidel:2021fxf}. Note that the remaining component $\lambda^{23}$ is completely unconstrained here. In fact, it parameterizes the residual internal U(1) freedom and will appear in the transformation of $\varphi$ below, so that it is only constrained by the boundary condition \eqref{eq:anglefalloff} to have a similar radial expansion, with fall-off
\be
\label{eq:lambdaexpansion}
\lambda^{23}=\O(1).
\ee 
The components of this expansion are otherwise arbitrary. What we finally get is that the tetrad residual symmetry generators can be parameterized by $\big(f(u,x^a), Y^a(x^b), \lambda^{23}(u,r,x^a)\big)$. 

Before computing the algebra of these parameters, we can ask if we have new transformation laws to write in the tetrad context. In fact, all the transformation laws computed previously are unchanged, and it only remains to compute the transformation law for the new solution space parameter $\varphi$. To get this, we can use the fact that 
\be
\delta\varphi=\f{1}{2}\epsilon^{ij}\big(\underline{E}_{ai}\delta\underline{E}_j^a-E_{ai}\delta E_j^a\big)=i\big(\underline{\bar{m}}_a\delta\underline{m}^a-\bar{m}_a\delta m^a\big).
\ee
We thus need the transformation law for the (real or complex) dyad and its reference. Looking at the right components in the tetrad transformation \eqref{full transformation of tetrad}, we get that
\be
\label{eq:transformationE}
\delta_{(f,Y,\lambda^{23})}E^a_i=\big(f\partial_u+\xi^r\partial_r+\pounds_{(Y+I)}\big)E^a_i+U^aE^b_i\partial_bf+\lambda^{23}{\epsilon_i}^jE^a_j.
\ee
To find the transformation law of the reference tetrad, one has to keep in mind that it is only a functional $\underline E_i[\gamma]$ of the angular metric; we can therefore use the transformation law \eqref{eq:gammatransfo} and obtain
\be
\label{eq:transformationEbar}
\delta_{(f,Y,\lambda^{23})}\underline{E}^a_i=\big(f\partial_u+\xi^r\partial_r+\pounds_{(Y+I)}\big)\underline{E}^a_i+U^a\underline{E}^b_i\partial_bf+\underline{E}^b_2\big(\pounds_{(Y+I)}\underline{E}_{3b}-\partial_bfU^c\underline{E}_{3c}\big){\epsilon_i}^j\underline{E}^a_j,
\ee
which is of course independent of $\lambda^{23}$.
Using \eqref{eq:transformationE} and \eqref{eq:transformationEbar} we finally obtain the transformation law
\be
\label{eq:phitransformation}
\delta_{(f,Y,\lambda^{23})}\varphi=\big(f\partial_u+\xi^r\partial_r+\pounds_{(Y+I)}\big)\varphi-\underline{E}^b_2\big(\pounds_{(Y+I)}\underline{E}_{3b}-\partial_bfU^c\underline{E}_{3c}\big)+\lambda^{23}.
\ee
We see here both that the angle $\varphi$ transform non-trivially under generators of the form $(f,Y,0)$, and that $(0,0,\lambda^{23})$ acts as an angle shift, hence the constraint \eqref{eq:lambdaexpansion} on $\lambda^{23}$'s expansion.

Let us now turn to the computation of the parameter algebra for residual symmetry generators of $\S_e^{\text{U(1)}}$. Using the corresponding modified bracket \eqref{eq:modifiedbracket} with field-independent parameters, we obtain that they form an algebra given by
\bsub
\be
Y_{12}&=[Y_1,Y_2],\\
f_{12}&=Y_1[f_2]-Y_2[f_1]+f_1\partial_uf_2-f_2\partial_uf_1,\\
\lambda^{23}_{12}&=\xi_1[\lambda^{23}_2]-\xi_2[\lambda^{23}_1]+\beta[\xi_1,\xi_2],
\ee
\esub
where $\xi_i=\xi(f_i,Y_i)$. In the first two lines, we recognize the previous metric algebra $\mathfrak{bmsw}_4$. It would have been satisfactory that, if turning off the residual U(1) generators $\lambda^{23}_i$, the metric algebra appeared as a subalgebra of this tetrad algebra, acting semi-directly on the residual U(1) factor. It is manifestly not the case because of the presence of the field-dependent cocycle
\be
\beta[\xi_1,\xi_2]=(\lambda_1^{03}\lambda_2^{12}-\lambda_2^{03}\lambda_1^{12})+(\lambda_1^{13}\lambda_2^{02}-\lambda_2^{13}\lambda_1^{02}),
\ee
where $\lambda^{IJ}_i=\lambda^{IJ}[\xi_i]$ according to the formulas in \eqref{eq:lambdas}. It represents the residual U(1) factor transformation produced by the commutator of two generators which themselves have a vanishing residual U(1) factor. 

What we face here precisely an instance of the phenomenon described in \ref{sec:obstruction}: solving the preservation of the tetrad gauge conditions gave a particular solution $\sigma[\xi]^{IJ}$, its coefficients being given by $\sigma[\xi]^{IJ}=\lambda^{IJ}[\xi]$ for $IJ\neq23$ and by $\sigma^{23}[\xi]=0$, from which the general solution can be written as $\sigma[\xi]^{IJ}+\delta^{[I}_2\delta^{J]}_3\lambda^{23}$. In terms of parameters the particular solution is $(f,Y,0)$, to which we can add any $(0,0,\lambda^{23})$, and the naive lift $(f,Y)\mapsto(f,Y,0)$ fails to realize the (parameter) metric asymptotic symmetry algebra as a subalgebra of the tetrad one. Let us now see how to cure this, first by hand using an educated guess from section \ref{sec:3}, then by using directly the shortcut this section provides.

\subsubsection{Cure by parameter redefinition}
We show here that we can reparameterize by hand the tetrad algebra in order to remove the cocycle above. More precisely, let us recall that when solving the preservation of the tetrad gauge conditions, part of the Lorentz freedom has been constrained through \eqref{eq:lambdas}, but the $23$ coefficient was left free except for its fall-off behavior. We have then interpreted $\lambda^{23}$ as the generator of residual U(1) transformations. What is natural to ask now is whether there exists a field-dependent redefinition $\lambda^{23}\mapsto\lambda^{23}[\xi]+\mu^{23}$ of the U(1) generator for which the cocycle vanishes. Equivalently, this amounts asking for a relevant prescription for a $\lambda^{23}[\xi]$ completing the formulas of \eqref{eq:lambdas}.

We have at our disposal here a reference tetrad and an associated comparison field, in the sense of section \ref{sec:referenceandcomparison}. Recall that the comparison field is the Lorentz U(1) rotation $\Lambda_0[\varphi]$ with angle $\varphi$. From the general result of section \ref{sec:3}, we know that a lift without cocycle is provided by asking that the comparison field is preserved under its action. What is natural to impose as a condition on $(\xi,\lambda)$, on top of the tetrad gauge and boundary conditions, is then that $\delta_{(\xi,\lambda)}\Lambda_0[\varphi]=0$, or equivalently that $\delta_{(\xi,\lambda)}\varphi=0$. Using the transformation law \eqref{eq:phitransformation} for $\varphi$, we see that this supplementary condition fixes $\lambda^{23}$ uniquely as a functional of $\xi$ by
\be
\label{eq:lambda23}
\lambda^{23}[\xi]=-\big(f\partial_u+\xi^r\partial_r+\pounds_{(Y+I)}\big)\varphi+\underline{E}^b_2\big(\pounds_{(Y+I)}\underline{E}_{3b}-\partial_bfU^c\underline{E}_{3c}\big),
\ee
which we can check to be $\O(1)$ as required.

We can then perform the reparameterization $\lambda^{23}\mapsto\lambda^{23}[\xi]+\mu^{23}$, the new free generator of the residual U(1) transformation being denoted by $\mu^{23}$. In terms of the new parameters $(f,Y,\mu^{23})$, the algebra becomes
\bsub
\be
Y_{12}&=[Y_1,Y_2],\\
f_{12}&=Y_1[f_2]-Y_2[f_1]+f_1\partial_uf_2-f_2\partial_uf_1,\\
\mu^{23}_{12}&=0.
\ee
\esub
If we denote by $\mathfrak{a}_\text{U(1)}$ this algebra and by $\mathfrak{l}_{\text{U(1)}}$ the ideal made of generators of the form $(0,0,\mu^{23})$, generating local U(1) transformations, then this time it is explicit that 
\be
\mathfrak{a}_\text{U(1)}=\mathfrak{bmsw}_4\oplus\mathfrak{l}_{\text{U(1)}}.
\ee
Note that because we work at the level of the parameterization, we take $\mu^{23}_i$ to be field-independent, so that there is no semi-direct action from the metric factor on the residual Lorentz one here. The only special property here is that the ideal $\mathfrak{l}_{\text{U(1)}}$ is abelian, which is why $\mu^{23}_{12}=0$. The result is then just as expected from section \ref{sec:structuretheorem}: the metric parameter algebra $\mathfrak{bmsw}_4$ is realized as a subalgebra of $\mathfrak{a}_\text{U(1)}$, through the Lie algebra homomorphic lift $(f,Y)\mapsto(f,Y,0)$, defined in the new parameterization.

\subsubsection{Use of the structure theorem}
Let us finally end by stating how we would have reached the same results, in a much quicker way, by direct use of the proposition \ref{prop:structureresult}. Indeed, an upshot of this result is that we can bypass the computations made above and find directly the right, cocycle-free lift of the metric asymptotic algebra. For this, the only thing we have to do is to compute the Lorentz transformation entering the action of metric asymptotic symmetries on the reference tetrad $\underline{e}$. That is, the Lorentz transformation $\rho_{\underline{e}}[\xi]$ defined by $\delta_\xi\underline{e}=\pounds_\xi\underline{e}+\rho_{\underline{e}}[\xi]\,\underline{e}$. Using the metric transformation laws \eqref{eq:expansionfreetransformation}, we obtain that this Lorentz transformation reads explicitly
\bsub
\label{eq:rho}
\be
\rho^{01}_{\underline{e}}&=-\partial_r\xi^r,\\
\rho^{0i}_{\underline{e}}&=e^{2B}\underline{E}^a_i\partial_af,\\
\rho^{1i}_{\underline{e}}&=\underline{E}^a_i\left(\partial_a\xi^r-\f{V}{2r}\partial_af\right),\\
\rho^{ij}_{\underline{e}}&=\underline{E}^b_2\big(\pounds_{(Y+I)}\underline{E}_{3b}-\partial_bfU^c\underline{E}_{3c}\big)\epsilon^{ij}.
\ee
\esub
Then, we know that the right compensating Lorentz transformation for a given $\xi$, that is the one adapted both to the tetrad gauge conditions and to the coordinate $\varphi$, is given in \eqref{eq:lift} by $s[\xi]=\text{Ad}_{\Lambda_0}\rho_{\underline{e}}[\xi]-\xi[\Lambda_0]\Lambda_0^{-1}$ that we can compute explicitly. The effect of the term with the adjoint action is to rotate the formula \eqref{eq:rho} by the angle $\varphi$, which amounts to trade every $\underline{E}_i$ for its rotated version $E_i$. With this, we recover the formulas in equation \eqref{eq:lambdas}. Then, the term in Lie derivative $-\xi[\Lambda_0]\Lambda_0^{-1}$ is precisely what brings the one of the form $-\pounds_\xi\varphi$ in the equation \eqref{eq:lambda23} for $\lambda^{23}[\xi]$. After that, we are done: to the obtained compensating Lorentz transformation, we can add an arbitrary local U(1) generator with parameter $\mu^{23}$, and all the results for the structure of the algebra we derived above follow naturally.

%%%%%%%%%%%%%%%%%%%%%%%%%%%%%%%%%%%%%%%%
%%%%%%%%%%%%%%%%%%%%%%%%%%%%%%%%%%%%%%%%

\section{Perspectives}
\label{sec:perspectives}
In this work, we have studied how the residual symmetries of a metric solution space can be realized in a first-order, tetrad description of that solution space, when part of the local Lorentz freedom is fixed by additional tetrad gauge conditions. We have shown that solving directly the preservation of these gauge conditions generically provides an embedding of the metric residual algebra in the tetrad one which fails to be a Lie algebra homomorphism, because of the presence of a field-dependent cocycle. We then obtained that having a reference field, that is a section of the tetrad solution space seen as a principal bundle over the metric one, always provides a lift of the metric residual algebra which is free of any cocycle. The resulting structure for the whole tetrad residual algebra as well as various direct upshots of this result have been discussed in section \ref{sec:summ}, then illustrated in detail through the example of section \ref{sec:BMSW-u-U(1)}. 

Let us finally give some perspectives and interesting directions in which this work could be used or pushed:
\begin{itemize}

\item \textbf{First-order charges, gyroscopic precession.} A direct continuation would be a detailed analysis of the charges that arise in the example of section \ref{sec:BMSW-u-U(1)}, for Cartan–Palatini Lagrangian but adding also Holst, Euler and Pontryagin terms. As discussed in section \ref{sec:summ}, the choice of lift only affects the charges through the residual Lorentz charges. These depend on the Lagrangian, on the residual group and on the choice of symplectic potential. With the standard potential, the Cartan–Palatini Lorentz charges vanish for these U(1) rotations, which are tangent to the corner. The Holst term, which underlies the so-called dual charges \cite{Godazgar:2020kqd}, gives them a charge whose variation is proportional to that of the area form of the corner, therefore it does not vanish when the boundary metric is unfixed, which is the case here. Finally, the Euler and Pontryagin terms can contribute as well \cite{Godazgar:2022foc}. This is therefore the natural setting in which the cocycle of a naive lift and the frame dependence of the charges should become visible. It also appears at the right place to discuss the link that may exist between these residual Lorentz charges and the precession of a gyroscope close to null infinity, as pointed out in \textit{e.g.} \cite{Seraj:2021rxd, Godazgar:2022pbx}. We plan to explore this in future work.

\item \textbf{Non-abelian examples.} The structure theorem obtained in section \ref{sec:3} holds for any residual subgroup $H$, which in particular does not need to be abelian. A natural generalization of the example studied in section \ref{sec:BMSW-u-U(1)} allowing for a non-abelian residual subgroup would be to drop the condition on $e^\mu_i$ to be tangent to the two-spheres. In that case, the only tetrad gauge condition would be $\ell=\partial_r$, and the residual subgroup would be the whole little group ISO(2) of the null radial generator, which is indeed non-abelian. In Newman--Penrose terms, this would amount to freeing the class I transformations, while keeping the U(1) rotation part of the class III transformations. Note that the reference tetrad used here would be a valid reference tetrad in that generalization, and it would only remain to parameterize conveniently the non-abelian residual freedom.

\item \textbf{Other boundaries.} The construction also does not depend on the nature of the asymptotics. It applies for instance to asymptotically (A)dS spacetimes, where it could be checked that the tetrad description of the $\Lambda$-BMS solution space inherits exactly the algebroid structure of the metric one \cite{Compere:2019bua}. It also applies to spatial infinity, where compensating Lorentz transformations have recently been used for the Holst action in \cite{Bakhoda:2026aeo} in a spirit that is close to that of the present paper, and finally to finite-distance boundaries and corners, see \textit{e.g.} \cite{Adami:2021nnf, Freidel:2020svx}.

\item \textbf{Use for fermions.} Fermions can only be coupled to gravity with a tetrad field. As they transform under the Lorentz part of the gauge generators, the action of metric asymptotic symmetries on fermionic fields, and hence the associated charges and their algebra, depends on the choice of compensating Lorentz transformations. The present construction provides a prescription for these compensating Lorentz transformations, once a reference frame is chosen. On isometries, we saw that it reduces to the Kosmann lift, which underlies the Lorentz--diffeomorphism charges used for black hole thermodynamics with fermions, for instance in \cite{Aneesh:2020fcr}. Away from isometries, it extends this prescription to asymptotic symmetries in a way that is compatible with their algebra, and it would be interesting to study the resulting fermionic charges.

\item \textbf{Other gauge theories, Carrollian gravity.} We think that the construction developed in section \ref{sec:3} could easily be generalized beyond tetrad gravity by adding other forms of internal gauge freedom than local Lorentz transformations. In particular, it could be interesting to see how it can be worked out in Einstein--Maxwell or Einstein--Yang--Mills theories. It could also be applied to first-order formulation of Carrollian gravity, see \textit{e.g.} \cite{Hartong:2015xda, Figueroa-OFarrill:2022mcy, Fiorucci:2025twa}, where Carrollian frames fibre over Carrollian structures as tetrads over metrics do. 

\end{itemize}

%%%%%%%%%%%%%%%%%%%%%%%%%%%%%%%%%%%%%%%%
%%%%%%%%%%%%%%%%%%%%%%%%%%%%%%%%%%%%%%%%

\section*{Acknowledgements}
The author greatly thanks his PhD advisor Marc Geiller, as well as Simone Speziale, Dima Fontaine and Yvann Gaudillot-Estrada, for all the fruitful discussions leading to the writing of the present paper.

\newpage

%%%%%%%%%%%%%%%%%%%%%%%%%%%%%%%%%%%%%%%%
%%%%%%%%%%%%%%%%%%%%%%%%%%%%%%%%%%%%%%%%

\appendix

\section{Global and local geometry}
\label{app:geometry}
The goal of this appendix is to give a brief account of the global differential geometry setting involved in first-order tetrad formalism and of how to go from the global to the local description, for both the fields and the covariance group acting on them. In particular, we recall the global origin of the gauge symmetry generators in tetrad formalism, which are pairs of a diffeomorphism and a local Lorentz generator in the local description. The derivation of this decomposition interestingly relies on the choice of a local section of a principal bundle, in a way that is closely analogous to the use of a reference field in section \ref{sec:3}. The usual local formulas used in local index form are then recovered. Finally, some remarks are made on covariant lifts of vector fields, among which the Kosmann lift and associated Kosmann Lie derivative. For general material on differential geometry, see \textit{e.g.} \cite{Kobayashi1963Foundations}. 

%%%%%%%%%%%%%%%%%%%%%%%%%%%%%%%%%%%%%%%%

\subsection{Fields: tetrads and connections}
In pure general relativity in metric formalism, the global geometric scene consists of a given spacetime differential manifold $\M$. Fields are then sections of the bundle Lor($\M$), \textit{i.e.} Lorentzian metrics on $\M$. For fields to exist, there are topological conditions on $\M$. Namely, a Lorentzian metric exists on any non-compact manifold, and on a compact one if and only if its Euler characteristic vanishes. What is the analogous story for the first-order tetrad formulation?

A way to present it starts the same: we are first given a spacetime differential manifold  $\M$. Let us then assume that we are also given of principal $\text{SO}^+(3,1)$-bundle $\P\xrightarrow{\pi}\M$. Consider then Minkowski spacetime $V\coloneqq(\mathbb{R}^4,\eta)$. The group $\text{SO}^+(3,1)$ acts on $V$ by Lorentzian isometries \textit{via} the fundamental representation $\rho$. With this, we build the associated ``Minkowski'' vector bundle $E=\P\times_\rho V$. Precisely because by definition $\text{SO}^+(3,1)$ preserves the Minkowski metric $\eta$ on $V$, the vector bundle $E$ is endowed with a global vector bundle metric $\eta_E$. Let us then define the bundle Iso($T\M,E$) whose fiber over $x\in\M$ consists of the set of all linear isomorphisms between $T_x\M$ and $E_x$. Tetrad fields are sections $e$ of this bundle. The second fields of interest in this context are finally given by the sections $\omega$ of the bundle Conn($\P$), \textit{i.e.} principal connections on $\P$. Using the identity component of the Lorentz group rather than the whole O(3,1) requires $\M$ to be orientable and time-orientable, and restricts to tetrads compatible with a chosen orientation and time orientation. Through such a tetrad, $\P$ is then identified with the bundle of oriented, time-oriented orthonormal frames of the induced metric defined right after.

The link between first- and second-order technology goes as follows. First, any tetrad field $e$ can be used to define a global Lorentzian metric $g$ on $T\M$ by
\be
\label{eq:induced metric}
g(X,Y)\coloneqq\eta_E(e(X),e(Y))\,,\qquad\forall X, Y \in \mathfrak{X}(\M).
\ee
Second, any principal connection $\omega$ on $\P$ defines a unique vector bundle connection $\nabla^E$ on $E$. Then, the tetrad $e$ can be used again to define a vector bundle connection on $T\M$ by
\be
\label{eq:tetrad postulate}
\nabla_XY\coloneqq e^{-1}(\nabla^E_{X}(e(Y)))\,,\qquad\forall X, Y \in \mathfrak{X}(\M).
\ee
This relation will be recognized below as the global version of the so-called tetrad postulate. The connection $\nabla$ on $T\M$ appears to be compatible with the metric $g$ defined by \eqref{eq:induced metric}. However, nothing imposes its torsion to vanish at this stage. It will only become the Levi-Civita connection associated with $g$ when going on shell of the vanishing torsion constraint coming from the Cartan--Palatini Lagrangian equations of motion.

%%%%%%%%%%%%%%%%%%%%%%%%%%%%%%%%%%%%%%%%

\subsection{Automorphisms: finite, infinitesimal and action on fields}
In pure general relativity in metric formalism, without any background or additional structure to be preserved, the automorphism group of the scene where fields ``take place'' is the group Diff($\M$) of spacetime diffeomorphisms, and it acts naturally by pullback on any given metric tensor and all the tensors constructed from it. At the infinitesimal level, the corresponding Lie algebra of generators is $\mathfrak{diff}(\M)=\mathfrak{X}(\M)$, acting on fields by Lie derivative.

In first-order formalism, the scene is somehow enlarged from $\M$ to the principal bundle $\P$. Even if fields in this context can still be seen as sections of bundles over $\M$, the group Diff($\M$) does not act naturally on these. What does act naturally this time is the automorphism group of the enlarged scene, that is the group Aut($\P$) of principal bundle automorphisms of $\P$. Let us recall that these are diffeomorphisms $\Phi : \P\rightarrow\P$, which commute with the right action of $\text{SO}^+(3,1)$ (so that they preserve the fibers). In that case, there is a unique covered spacetime diffeomorphism $\varphi\in\text{Diff}(\M)$ such that $\varphi\circ\pi=\pi\circ\Phi$. How does such an automorphism then act on the fields?

First, a given $\Phi\in\text{Aut}(\P)$ induces a unique vector bundle automorphism $\Phi_E$ of the associated bundle $E$. The natural induced action of $\Phi$ on a tetrad field is then $\Phi^*e\coloneqq\Phi_E^{-1}\circ e\circ T\varphi$, which is of course again a valid tetrad field. For a principal connection field $\omega$, the action is simply given by the pullback $\Phi^*\omega$ of $\omega$, as a one-form on $\P$.

At the infinitesimal level, the Lie algebra of generators of automorphisms is $\mathfrak{aut}(\P)=\mathfrak{X}_G(\P)$, formed by $\text{SO}^+(3,1)$-invariant vector fields on $\P$. As in the finite case, an infinitesimal automorphism $X\in\mathfrak{aut}(\P)$ can be projected to a spacetime vector field $\xi=\pi_*X\in\mathfrak{diff}(\M)$. Without additional structure, there is however no canonical lift allowing one to define an automorphism generator from a spacetime vector field. What is the corresponding action on the fields?

Mirroring the finite case, $X\in\mathfrak{aut}(\P)$ induces a unique infinitesimal vector bundle automorphism generator $X_E$. The action of $X$ on a tetrad field is then defined by $\delta_Xe=X_E\circ e-e\circ\pounds_\xi$. For a principal connection field $\omega$, the infinitesimal action is simply the Lie derivative $\delta_X\omega=\pounds_X\omega$.

%%%%%%%%%%%%%%%%%%%%%%%%%%%%%%%%%%%%%%%%

\subsection{From global to local: recovering indices}
The passage to the usual local description in the metric case requires a choice of a coordinate chart on $\M$ over a coordinate patch $U\subset\M$, giving coordinates $x^\mu$ and its index form $g_{\mu\nu}$ to the metric field.

In the first-order case, a supplementary choice has to be made in order to also get the internal Lorentz index and the local automorphism expressions. This choice is that of a local trivialization $\P_{\vert U}\xrightarrow{\sim} U\times\, \text{SO}^+(3,1)$ over the coordinate patch $U$\footnote{Note that if $U$ is contractible, trivializability is guaranteed.}. Let us recall the important fact that for principal bundles, local trivializations and local sections are in one-to-one correspondence. Indeed, let $s : U \rightarrow \P$ be a local section. A trivialization is then given by $t_s : p\in\P\mapsto(\pi(p),g_s(p))\in U\times \text{SO}^+(3,1)$, where $g_s(p)$ is the unique Lorentz group element such that $p=s(x)\cdot g_s(p)$, the centered dot denoting the free and transitive right action of the Lorentz group on $\P$, part of its definition as a principal bundle. The equivariance of this trivialization is obvious. One can conversely show that a given local trivialization uniquely determines a local section.

One can now read the local expression of a given automorphism $\Phi$ in the local trivialization. On a couple $(x,g)\in U\times \text{SO}^+(3,1)$, the action of the automorphism is uniquely written as $\Phi(x,g)=(\varphi(x),\gamma(x)\,g)$\footnote{Technically, we need here to assume that $\varphi(x)$ is also in the local patch $U$.}, where $\varphi$ is the covered diffeomorphism and $\gamma : U\rightarrow \text{SO}^+(3,1)$ is a Lorentz group valued function on $U$. At the infinitesimal level, the couple $(\varphi,\gamma)$ encoding the local action of $\Phi$ then becomes a couple $(\xi,\lambda)$ of a local diffeomorphism generator on $U$ and a local Lorentz generator $\lambda : U \rightarrow \mathfrak{so}(3,1)$\footnote{Of course, the couple $(\xi,\lambda)$ is the infinitesimal version of $(\varphi,\gamma)$ only if $\Phi$ is connected to the identity.}. It is then easy to show that the algebra of such infinitesimal generators is given by
\be
\label{eq:algebraofgenerators}
[(\xi_1,\lambda_1),(\xi_2,\lambda_2)]=([\xi_1,\xi_2],\xi_1[\lambda_2]-\xi_2[\lambda_1]-[\lambda_1,\lambda_2]),
\ee
where the bracket of vector fields and Lorentz generators are the usual ones and $\xi[\lambda]$ is the directional derivative of $\lambda$ along $\xi$. Note that the Lorentz part $\gamma$ of $\Phi$ acts on $\P$ by left multiplication, and its pullback acts on the components of the fields by $\gamma^{-1}$, which defines a right action of the gauge group on fields. In the main text, we use instead the left action $e\mapsto\Lambda e$, with $\Lambda=\gamma^{-1}$. At the infinitesimal level, this amounts to redefining $\lambda\mapsto-\lambda$ which flips the sign of the Lorentz bracket in \eqref{eq:algebraofgenerators} such that we recover the bracket \eqref{eq:generatorbracket} of the main text. The structure of the algebra does not depend on this convention: elements of the form $(\xi,0)$ form a subalgebra, isomorphic to $\mathfrak{diff}(U)$, and ``vertical'' or pure Lorentz generators, of the form $(0,\lambda)$, form an ideal. This shows that we locally have the semi-direct sum structure
\be
\mathfrak{aut}(\P_{\vert U})=\mathfrak{diff}(U)\rsemidirsum\C^\infty(U,\mathfrak{so}(3,1)).
\ee

Let us now use the local trivialization to recover the usual local index formulas used in tetrad formalism. First, the local trivialization of $\P$ induces a local trivialization of the associated bundle $E_{\vert U}\xrightarrow{\sim} U\times\,V$\footnote{Indeed, elements of $E$ are equivalence classes $[p,v]$ of pairs $(p,v)\in\P\times V$. The section associated with the trivialization of $\P$ gives a canonical representative for each equivalence class: the one whose first element belongs to the image of the section.}. We can now endow the model fiber $V$, which is Minkowski space, with a fixed basis $(e_I)$. With the coordinate chart $x^\mu$, we similarly get a coordinate basis $(\partial_\mu)$ on the model fiber of $TU\simeq U\times\mathbb{R}^4$. Finally, we can assign to any given tetrad field $e$ its local expression $e_\mu^I$, defined by $e(\partial_\mu)=e_\mu^Ie_I$ and similarly for the inverse tetrad $e^\mu_I$. Now, for a given connection field $\tilde\omega$ which we recall to be initially a form on $\P$, we can use the trivialization's section to pull it back and define the local connection form $\omega=s^*\tilde\omega\in\Omega^1(U,\mathfrak{so}(3,1))$. Using the local coordinate co-basis as well as the conventional basis $J_{IJ}$ with two internal indices for $\mathfrak{so}(3,1)$, the index form ${\omega_\mu}^{IJ}$ is defined by $\omega=\f{1}{2}{\omega_\mu}^{IJ}J_{IJ}\d x^\mu $. We can now translate the equations \eqref{eq:induced metric} and \eqref{eq:tetrad postulate} to their usual local form
\be
g_{\mu\nu}&=\eta_{IJ}e_\mu^Ie_\nu^J,\\
\partial_\mu e_\nu^I&=\Gamma^\rho_{\mu\nu}e^I_\rho-{{\omega_\mu}^I}_Je^J_\nu,
\ee
the first one defining the local tangent space metric induced by $e$ and $\eta$, the second defining the local tangent space connection induced by $e$ and $\omega$. Generally, we can now use the tetrad and its inverse to trade spacetime greek indices for latin internal indices, and vice-versa, as well as use the spacetime and internal metric to raise and lower these indices. Let us finally write the important formulas of the action of automorphism generators on fields. First, we can use the constructed coordinates to give indices to the generators: $(\xi,\lambda)=(\xi^\mu\partial_\mu,\f{1}{2}\lambda^{IJ}J_{IJ})$. Then, on fields, we have
\be
\delta_{(\xi,\lambda)}e_\mu^I&=\pounds_\xi e_\mu^I+{\lambda^I}_Je^J_\mu,\\
\delta_{(\xi,\lambda)}{\omega_\mu}^{IJ}&=\pounds_\xi{\omega_\mu}^{IJ}-\partial_\mu\lambda^{IJ}-[\omega_\mu,\lambda]^{IJ},
\ee
which are the formulas of equation \eqref{eq:transformationlaws}.

%%%%%%%%%%%%%%%%%%%%%%%%%%%%%%%%%%%%%%%%

\subsection{Remarks on the Kosmann lift}
\label{app:kos}
About the possibility of lifting a spacetime vector field to a bundle automorphism generator, let us say once again that we succeed in doing it homomorphically only locally. Importantly, the form of the lifted vector $(\xi,0)$ is not invariant under a change of trivialization. Indeed a change of trivialization corresponds locally to the action of a vertical automorphism, and under an infinitesimal one, we have $\delta_{(0,\lambda)}(\xi,0)=(0,-\xi[\lambda])$: it turned on a vertical component! This lack of covariance under a change of trivialization is mirrored in the formulas for the action of a lifted diffeomorphism, that is simply the Lie derivative, on the fields
\be
\label{eq:Liederivativeaction}
\delta_{(\xi,0)}e^I&=\pounds_\xi e^I=(\iota_\xi \d_\omega e^I+ \d_\omega \iota_\xi e^I)-(\iota_\xi{\omega^I}_J)e^J,\\
\delta_{(\xi,0)}\omega^{IJ}&=\pounds_\xi \omega^{IJ}=\iota_\xi F^{IJ}+\d_\omega(\iota_\xi\omega^{IJ}),
\ee
where $F^{IJ}$ is the curvature of $\omega^{IJ}$ and where we can see for both that the second term in the right-hand side is not Lorentz covariant. 

We can then wonder what would be a globally well-defined lift, whose local formula is actually covariant. In fact, if we allow ourselves to use the connection in its definition, there is a canonical associated horizontal lift, given by $\xi\mapsto\hat{\xi}_\omega=(\xi,\iota_\xi\omega)$. It is then easy to check that the formulas in \eqref{eq:Liederivativeaction}, with $\hat{\xi}_\omega$ instead of $(\xi,0)$, lose their non-covariant term. This lift is however not Lie homomorphic as we have
\be
[\hat{\xi}_{\omega,1},\hat{\xi}_{\omega,2}]=\widehat{[\xi_1,\xi_2]}_\omega+(0,\iota_{\xi_1}\iota_{\xi_2}F).
\ee
We see that for this to define a global Lie homomorphic lift, $\omega$ has to be flat. The existence of a global flat connection is actually a very restrictive topological condition of $\P$, which is usually not satisfied. 

If we allow ourselves to use, in addition to a connection $\omega$, a tetrad field $e$, then there is another lift of interest. Indeed, first note that we can add to $\hat{\xi}_\omega$ any vertical piece of the form $(0,\lambda_\xi)$. Taking then $\omega=\omega[e]$ the Levi-Civita spin connection associated with $e$, the Kosmann lift $\K^e_\xi$ is defined by the choice $\lambda_\xi^{IJ}=\nabla^{[I}\xi^{J]}$, where $\nabla$ stands for the Levi-Civita connection for the metric induced by $e$, in Lorentz indices. Explicitly, we have
\be
\K^e_\xi\coloneqq(\xi,\,\iota_\xi\omega^{IJ}[e]+\nabla^{[I}\xi^{J]})=(\xi,\,e^{\mu[I}\pounds_\xi e^{J]}_\mu),
\ee
which depends in fact only on the tetrad. Again, this lift is global but not Lie homomorphic in general. Its nice property is its compatibility with spacetime isometries. Indeed, we have that
\be
\delta_{\K^e_\xi}e^I_\mu=\f{1}{2}e^{I\nu}\pounds_\xi g_{\mu\nu}[e],
\ee
so that $\delta_{\K_\xi^e}e=0\Leftrightarrow\pounds_\xi g[e]=0$. As argued in section \ref{sec:structuretheorem}, this property is also satisfied by the lift constructed in that section, which coincides with the Kosmann lift on spacetime isometry generators.  See \textit{e.g.} \cite{Jacobson:2015uqa, Prabhu:2015vua, DePaoli:2018erh, Aneesh:2020fcr} for discussions involving these objects on questions related to the symplectic structure in first-order formalism, black hole thermodynamics and their use in the presence of fermions.

Let us stress, however, that neither the horizontal lift associated with a connection nor the Kosmann lift is relevant for the problem addressed in the main text. Both fail in general to preserve tetrad gauge conditions and to realize the metric diffeomorphism algebra in that of automorphism generators, in particular the subalgebra of metric asymptotic symmetries.

%%%%%%%%%%%%%%%%%%%%%%%%%%%%%%%%%%%%%%%%

\section{Complements on the example}
\label{app:complements}
\subsection{Evolution equations}
\label{app:evolutionequation}
Using the notation of section \ref{sec:metricsolutionspace}, we write here the evolution equations for the mass and angular momentum aspects $M$ and $N^a$ of the metric solution space. Let us first introduce
\be
\gamma_0\coloneqq\f{1}{4}\partial_u\ln\sqrt{q},\q N_{ab}\coloneqq\partial_uC_{ab},\q\widetilde{C}^{ab}\coloneqq{\varepsilon^a}_cC^{cb}
\ee
with $\varepsilon_{ab}\coloneqq\sqrt{q}\,\epsilon_{ab}$, $\epsilon_{ab}$ being the Levi-Civita symbol\footnote{More generally, we define the dual of a tensor by $\widetilde{T}^{a_1... a_n}={\varepsilon^{a_1}}_bT^{ba_2... a_n}$.}, as well as
\bsub
\be
\M&\coloneqq M+\f{1}{8}C^{ab}N_{ab}-\f{1}{4}\gamma_0[CC],\\
\widetilde{\M}&\coloneqq\f{1}{8}\big(2D_aD_b-N_{ab}\big)\widetilde{C}^{ab},\\
\M_{ab}&\coloneqq q_{ab}\M+\epsilon_{ab}\widetilde{\M},\\
\P_a&\coloneqq-\f{3}{2}N_a+\f{3}{32}\partial_a[CC]+\f{3}{4}C_{ab}D_cC^{bc},\\
\J_a&\coloneqq\f{1}{4}\big(2D^bN_{ab}+\partial_aR\big)-D^b\big(\gamma_0C_{ab}\big),\\
\N_{ab}&\coloneqq\f{1}{2}\partial_uN_{ab}-\partial_u\big(\gamma_0C_{ab}\big)-2D_{\la a}\partial_{b\ra}\gamma_0.\label{curly N}
\ee
\esub
These definitions are motivated because they are the right spin-weighted quantities encoding the leading orders of the Newman--Penrose Weyl scalars, see \textit{e.g.} \cite{Freidel:2021fxf, Geiller:2024bgf}. Here, they serve to write neat formulas for the evolution equation for the mass and angular momentum. They take the form
\bsub
\be
\big(\partial_u+6\gamma_0\big)\M&=\f{1}{2}D_a\J^a+\f{1}{4}C^{ab}\N_{ab},\label{metric EOM M}\\
\big(\partial_u+4\gamma_0\big)\P_a&=D^b\M_{ab}+C_{ab}\J^b.\label{metric EOM Pd}
\ee
\esub
Note that we can also write an evolution equation for the dual mass $\widetilde{\M}$ which mirrors that for $\M$:
\be
\big(\partial_u+6\gamma_0\big)\widetilde{\M}=\f{1}{2}D_a\widetilde{\J}^a+\f{1}{4}C^{ab}\widetilde{\N}_{ab}.\label{metric EOM M dual}
\ee
However, this equation is tautological and simply follows from the previous definitions. For the evolution equations of the subleading angular aspects $E_{ab}^n$, see \textit{e.g.} \cite{Geiller:2024bgf}.

%%%%%%%%%%%%%%%%%%%%%%%%%%%%%%%%%%%%%%%%

\subsection{Expansion of the metric transformation laws}
\label{app:transformationlaws}
From the radial expansion of the formulas of \eqref{eq:expansionfreetransformation} and with the definitions of appendix \ref{app:evolutionequation}, one finds the more familiar transformation laws for various metric solution space parameters:
\bsub
\be
%%%%%%%%%%%%
\delta_\xi\ln\sqrt{q}&=f\partial_u\ln\sqrt{q}+D_aY^a-2W,\\
%%%%%%%%%%%%
\delta_\xi q_{ab}&=\big(f\partial_u+\pounds_Y-2W\big)q_{ab},\\
%%%%%%%%%%%%
\delta_\xi C_{ab}&=\big(f\partial_u+\pounds_Y-W\big)C_{ab}-2D_{\langle a}\partial_{b\rangle}f,\\
%%%%%%%%%%%%
\delta_\xi N_{ab}&=\big(f\partial_u+\pounds_Y\big)N_{ab}-2\partial_u\big(D_{\la a}\partial_{b\ra}f\big)-\partial_uWC_{ab},\\
%%%%%%%%%%%%
\delta_\xi\M&=\big(f\partial_u+\pounds_Y+3W\big)\M+\J^a\partial_af,\\
%%%%%%%%%%%%
\delta_\xi\widetilde{\M}&=\big(f\partial_u+\pounds_Y+3W\big)\widetilde{\M}+\widetilde{\J}^a\partial_af,\\
%%%%%%%%%%%%
\delta_\xi\P_a&=\big(f\partial_u+\pounds_Y+2W\big)\P_a+3\M_{ab}\partial^bf.
%%%%%%%%%%%%
\ee
\esub
For the transformation laws of the subleading angular aspects $E_{ab}^n$, see \textit{e.g.} \cite{Geiller:2024bgf}.

%%%%%%%%%%%%%%%%%%%%%%%%%%%%%%%%%%%%%%%%

\subsection{Expression of the spin connection}
\label{app:spinconnection}
In order to write explicitly the components of the spin connection of the tetrad solution space, we introduce
\bsub
\be
B_a&\coloneqq r\partial_aB+\f{r}{2}e^{-2B}\gamma_{ab}\partial_rU^b,\\
W^a&\coloneqq r^2e^{-2B}U^a,\\
X_{ab}&\coloneqq\f{1}{2r}\partial_r\gamma_{ab},\\
K_{ab}&\coloneqq\partial_u\gamma_{ab}+2\mathcal{D}_{(a}U_{b)},\\
\omega^{ij}_a&\coloneqq E^{bi}\mathcal{D}_aE^j_b.\label{2d omega connection}
\ee
With this we find
\be
%%%%%%%%%%%%%%%%%%%%
\omega^{01}
&=-\left(\f{1}{2}\partial_r\left(\f{V}{r}\right)+\f{1}{r}\big(V\partial_rB+B_aU^a\big)\right)\de u+2\partial_rB\de r+\f{1}{r}B_a\de x^a,\\
%%%%%%%%%%%%%%%%%%%%
\omega^{0i}&=\f{1}{r}e^{2B}\Big(B_a+X_{ab}W^b-2r\partial_aB\Big)E^{ai}\de u-rX_{ab}E^{ai}\de x^b,\\
%%%%%%%%%%%%%%%%%%%%
\omega^{1i}
&=\f{1}{2r^2}\Big(r\partial_aV+VB_a+\big(VX_{ab}+K_{ab}\big)W^b\Big)E^{ai}\de u-\f{1}{r}E^{ai}B_a\de r-\f{1}{2}e^{-2B}\big(VX_{ab}+K_{ab}\big)E^{ai}\de x^b,\\
%%%%%%%%%%%%%%%%%%%%
\omega^{ij}&=\Big(E^{a[i}\partial_uE^{j]}_a+E^{a[i}E^{j]b}\mathcal{D}_aU_b\Big)\de u+E^{a[i}\partial_rE^{j]}_a\de r+\omega^{ij}_a\de x^a,
%%%%%%%%%%%%%%%%%%%%
\ee
\esub
recovering the formulas of \cite{Freidel:2021fxf, Godazgar:2022foc}.

%%%%%%%%%%%%%%%%%%%%%%%%%%%%%%%%%%%%%%%%

\subsection{Alternative for the reference dyad}
\label{app:alternativedyad}
In \eqref{eq:referencem}, we made a choice for the reference dyad $\underline{m}^a$ for the angular metric $\gamma_{ab}$, namely
\be
\label{eq:referencem2}
\underline{m}=\sqrt{\f{\gamma_{\theta\theta}}{2\gamma}}\left(\f{\sqrt{\gamma}+i\gamma_{\theta\phi}}{\gamma_{\theta\theta}}\,\partial_\theta-i\partial_\phi\right),
\ee
the associated real reference dyad $\underline{E}_i=(\underline{E}_2,\underline{E}_3)$ being given by
\be
\underline{E}_2\coloneqq\f{1}{\sqrt{2}}(\underline{\bar{m}}+\underline{m})=\f{1}{\sqrt{\gamma_{\theta\theta}}}\partial_\theta,
\q\q
\underline{E}_3\coloneqq\f{1}{i\sqrt{2}}(\underline{\bar{m}}-\underline{m})=-\sqrt{\f{\gamma_{\theta\theta}}{\gamma}}\left(\f{\gamma_{\theta\phi}}{\gamma_{\theta\theta}}\,\partial_\theta-\partial_\phi\right).
\ee

The advantage of this reference is that it is written as a closed expression and does not need any radial expansion. From the point of view of Newman--Penrose formalism, it is however not convenient since the associated spin coefficient $\epsilon$ is non-vanishing, and given by
\be
\label{epsilon}
\epsilon=\f{1}{2}\Big(\underline{\bar{m}}^\mu\nabla_{(\ell)}\underline{m}_\mu-n^\mu\nabla_{(\ell)}\ell_\mu\Big)=\partial_rB-\f{i}{4}\text{Im}\big(\partial_r\gamma_{ab}\underline{m}^a\underline{m}^b\big).
\ee
In particular, this implies that the expansion of the complex dyad, if needed, is ``complicated''. If we write this expansion as $\underline{m}=\sum_{n\ge1}\underline{m}_nr^{-n}$, then the first coefficients are given by
\be
\underline{m}_1=\sqrt{\f{q_{\theta\theta}}{2q}}\left(\f{\sqrt{q}+iq_{\theta\phi}}{q_{\theta\theta}}\,\partial_\theta-i\partial_\phi\right),
\q
\underline{m}_2=-(\sigma_2\bar{\underline m}_1+2\epsilon_2\underline m_1).
\ee
Similarly, the expansion of the real dyad is ``complicated'' and, in particular, not symmetric between $i=2$ and $i=3$. Explicitly, it takes the form
\bsub
\be
\underline{E}^a_2&=\f{1}{r}\hat{\underline{E}}^a_2-\f{1}{2r^2}\hat{\underline{E}}^a_2\big(C_{bc}\hat{\underline{E}}^b_2\hat{\underline{E}}^c_2\big)+\mathcal{O}(r^{-3}),\\
\underline{E}^a_3&=\f{1}{r}\hat{\underline{E}}^a_3+\f{1}{2r^2}\Big(\hat{\underline{E}}^a_3\big(C_{bc}\hat{\underline{E}}^b_3\hat{\underline{E}}^c_3\big)-2C^a_b\hat{\underline{E}}^b_3\Big)+\mathcal{O}(r^{-3}),
\ee
\esub
where the leading order vectors are
\be\label{leading real dyad}
\hat{\underline{E}}_2=\f{1}{\sqrt{q_{\theta\theta}}}\partial_\theta,
\q\q
\hat{\underline{E}}_3=-\sqrt{\f{q_{\theta\theta}}{q}}\left(\f{q_{\theta\phi}}{q_{\theta\theta}}\,\partial_\theta-\partial_\phi\right).
\ee

Alternatively, one can perform a local U(1) rotation of the reference dyad \eqref{eq:referencem2} given by
\be
\underline{m}\mapsto\underline{m}'\coloneqq\exp\left(2i\int^r\text{Im}(\epsilon)\right)\underline{m}.
\ee
Under this rotation, spin coefficients transforms and in particular the new $\epsilon'$ has no imaginary part (see \cite{Geiller:2024bgf}). The expansion of the new dyad simplifies and becomes
\be
\underline m'_1=\underline m_1=\sqrt{\f{q_{\theta\theta}}{2q}}\left(\f{\sqrt{q}+iq_{\theta\phi}}{q_{\theta\theta}}\,\partial_\theta-i\partial_\phi\right),
\quad\quad
\underline m'_2=-\sigma_2\bar{\underline m}_1,
\quad\quad
\underline m'_3=\f{1}{2}\sigma_2\bar{\sigma}_2\underline m_1.
\ee
Similarly, the expansion of the real dyad becomes simple and symmetrical in the indices $i$. It reads
\be\label{rotated real frame}
\underline E'^a_i=\f{1}{r}\hat{\underline E}^a_i-\f{1}{2r^2}C^a_b\hat{\underline E}^b_i+\f{1}{16r^3}[CC]\hat{\underline E}^a_i+\mathcal{O}(r^{-4}),
\ee
with $\hat{\underline E}^a_i$ given once again by \eqref{leading real dyad}. Again, the reference tetrad obtained by using the transformed reference dyad is perfectly valid and is useful for some computations, but it does not provide closed formulas.

%%%%%%%%%%%%%%%%%%%%%%%%%%%%%%%%%%%%%%%%
%%%%%%%%%%%%%%%%%%%%%%%%%%%%%%%%%%%%%%%%

\newpage
\phantomsection
\addcontentsline{toc}{section}{References}
\bibliographystyle{Biblio}
\bibliography{Biblio}

\end{document}